\documentclass[twocolumn]{aastex701}
\usepackage{amsmath}
\usepackage{xspace}
\usepackage{CJK}

\newcommand{\model}{\texttt{CausticFlow}\xspace}

\begin{document}
\begin{CJK*}{UTF8}{gbsn}
\title{CausticFlow: An Efficient Machine Learning Framework Combining Neural Differential Equations \& Normalizing Flows for Binary Microlensing Parameter Inference}

\author[orcid=0009-0000-7721-6342,gname='Haibin',sname='Ren']{Haibin Ren (任海滨)}
\affiliation{Department of Astronomy, Tsinghua University, Beijing 100084, China}
\email[show]{rhb23@mails.tsinghua.edu.cn}

\author[orcid=0000-0003-4027-4711,gname='Wei',sname='Zhu']{Wei Zhu (祝伟)}
\affiliation{Department of Astronomy, Tsinghua University, Beijing 100084, China}
\email[show]{weizhu@mail.tsinghua.edu.cn}

\correspondingauthor{Haibin Ren, Wei Zhu}
\begin{abstract}

We introduce \model, a machine learning framework that combines neural controlled differential equations with normalizing flows to infer binary microlensing parameters. This architecture naturally handles irregularly sampled time series and data gaps while flexibly capturing strongly correlated and multimodal posterior distributions. Trained on simulated KMTNet-like light curves, \model generates posterior samples in a fraction of a second, with maximum-a-posteriori estimates achieving typical precisions of $\sim17\%$ for the mass ratio $q$ and $\sim3\%$ for the projected separation $s$. When used as a proposal distribution for downstream local optimization, the framework improves these precisions to $<5\%$ and $<1\%$, respectively, and recovers model $\chi^2$ for $\sim80\%$ of simulated events. We test the generalizability of the framework on 10 real binary lensing events characterized by higher-order effects, varied cadences, and real-world noise. Despite these mismatches between simulation and reality, \model successfully recovers the model parameters, light-curve morphology, and lensing geometry for 7 of the 10 events after simple local refinement, achieving precision levels comparable to those found for simulated data in 10 CPU minutes per event. These results demonstrate that \model acts as a fast and robust proposal engine that provides informative starting points for downstream local optimization, bridging the gap between the rapid influx of data and the need for systematic modeling in large-scale microlensing surveys such as Roman, CSST, and ET.

\end{abstract}

\section{Introduction}
Gravitational microlensing has become a powerful method for detecting and characterizing dark or faint objects, including exoplanets and stellar binaries \citep{mao1991,Gould&Loeb1992}.
Over the past two decades, ground-based surveys such as the Optical Gravitational Lensing Experiment (OGLE, \citealt{OGLEIV}), the Microlensing Observations in Astrophysics survey (MOA, \citealt{Sako2008}), and the Korea Microlensing Telescope Network (KMTNet, \citealt{kmtnet2016}) have accumulated a large archive of microlensing light curves.
Upcoming space-based facilities, including the Nancy Grace Roman Space Telescope \citep{spergel2015wide,penny2019} the Earth 2.0 Microlensing Telescope \citep{CMST,Ge2022}, and the Chinese Space Station Survey Telescope \citep{CSST2026, Yan2022}, are expected to further increase the number of detected events.
These developments make systematic and scalable analysis of microlensing events increasingly important.

Among all microlensing events, binary lens events, including those produced by both planetary and stellar binary lenses, are of particular interest because they often exhibit rich anomaly structures and, in many cases, finite-source effects associated with caustic approaches or crossings \citep{witt&mao1994,zhu2014,Jung2022}.
Such effects enable measurements of the angular Einstein radius, $\theta_{\mathrm{E}}$, which, when combined with high-resolution imaging \citep{bennett2007,Terry2026_HST_Bulge_Survey} or microlensing parallax measurements \citep{Gould1992,Gould2000}, can lead to mass measurements for large samples of lenses.
A systematic analysis of binary lens events therefore offers a promising route to studying the cold-planet population \citep{Wise,Suzuki2016,Zang2025}, the binarity and initial mass function (IMF) in the Galactic bulge \citep{CalchiNovati2008,Wegg2017,mroz2017}, and the population of stellar remnants.

At present, systematic analyses of stellar binary lens events remain limited to selected subsets, such as brown-dwarf candidates (e.g., \citealt{Han2026}) or events with well-separated bumps \citep{Oliveira2025}.
This limitation has two main causes.
First, valuable stellar binary lens systems are often difficult to distinguish from the much larger binary lens population using light-curve morphology alone.
By contrast, planetary events can often be identified through short-duration anomalies superposed on an otherwise single-lens light curve \citep[e.g.,][]{Zang2021}.
For systematic studies of the stellar binary lens population, meaningful sample selection therefore becomes possible only after detailed light curve modeling.
Second, stellar binary lens modeling is computationally expensive, making exhaustive modeling of all events almost prohibitive. 
The parameter space contains strong correlations and multiple discrete degeneracies induced by caustic features, including the well-known close/wide degeneracy \citep{Griest1998, Dominik1999, an2005} and many more \citep{gaudi&gould1997,Han2006,zhang2022ubiquitous}.
While planetary events often admit good initial guesses from heuristics \citep{Gould&Loeb1992,gaudi&gould1997,Han2006,Gaudi2012,Zhang&Gaudi2022}, the initial guesses for stellar binary lens events are much more complicated \citep{Liebig2015,bozza2024}.
The traditional modeling strategy combines brute-force searches to locate possible solutions and Markov chain Monte Carlo (MCMC) sampling to explore the local posterior around each solution.
Although this remains essential for detailed modeling of individual events of interest, its computational cost limits the statistical power of large-scale microlensing analyses.

Machine learning provides a promising route toward scalable inference for large microlensing datasets.
Simulation-based inference (SBI), also known as likelihood-free inference, is well suited to problems for which forward simulations are available but direct likelihood evaluation or posterior exploration is expensive \citep{cranmer2020frontier,deistler2025simulation}.
In particular, neural posterior estimation (NPE) amortizes Bayesian inference by training a conditional density estimator to learn the mapping from observed data to the corresponding posterior distribution \citep{papamakarios2016fast,lueckmann2017flexible,greenberg2019automatic}.
Once trained, such a model can rapidly generate learned posterior samples for new light curves without repeating a full global search.
NPE-based approaches have already been explored for microlensing inference \citep{zhang2021romanML,zhao&zhu2022,Smyth2025_Transformer_Embeddings}.
For example, \citet{zhang2021romanML} combined a ResNet--GRU embedding network with normalizing flows as a flexible conditional density estimator, focusing on a fairly ideal dataset with high-quality and regularly sampled light curves.
By contrast, \citet{zhao&zhu2022} used Neural Controlled Differential Equations (Neural CDEs) to handle more realistic light curves with noisy measurements and irregular sampling, but they modeled the posterior with a Gaussian mixture model, which limits flexibility for strongly multimodal and pathological posteriors.

In this work, we present \model, a neural posterior estimation framework designed to handle irregular, noisy microlensing light curves while preserving flexible posterior structure.
\model combines Neural CDEs, which provide a natural representation of irregularly sampled time series, with normalizing flows, which provide an expressive conditional density estimator for correlated and multimodal posteriors.
We train and validate the model on simulated binary lens light curves, assess its posterior calibration, compare representative cases against MCMC reference posteriors, and use the learned posterior as a proposal distribution for downstream optimization.
We then apply the workflow to real KMTNet events to test its robustness under realistic noise, incomplete coverage, and model mismatch.

The remainder of this paper is organized as follows.
Section~\ref{sec:overview} introduces the \model architecture.
Section~\ref{sec:data_generation_and_training} describes the simulated dataset and training procedure.
Section~\ref{sec:simulation_results} presents the validation on simulated data.
Section~\ref{sec:real_events} applies the workflow to real KMTNet events.
Section~\ref{sec:discussion} discusses the implications, limitations, and future extensions of the framework.

\section{Overview of \model}\label{sec:overview}

\begin{figure*}[htb!]
	\centering
	\includegraphics[width=\linewidth]{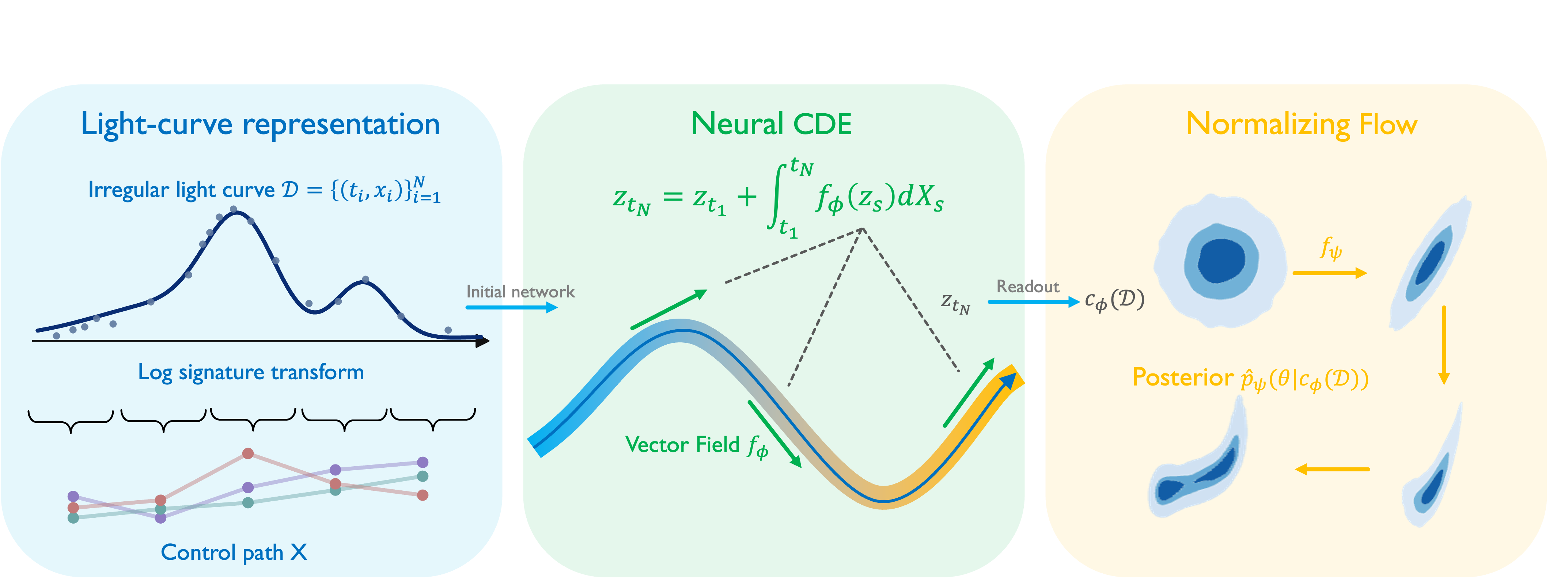}
	\caption{Overview of the \model architecture.
        Given an irregularly sampled light curve $\mathcal{D}=\{(t_i,x_i)\}$, the framework first applies a log-signature transformation to construct a compact control path $X$.
        An initial network maps the initial value of the control path to the initial hidden state of the Neural CDE. 
        The Neural CDE then evolves this hidden state along the full control path and produces the final summary state $z_{t_N}$.
        A readout network maps $z_{t_N}$ to the conditioning vector $c_\phi(\mathcal{D})$, which is passed to a conditional normalizing flow.
        The flow transforms a simple base distribution into a surrogate posterior $p(\theta \mid c_\phi(\mathcal{D}))$ over the predicted binary lens parameters defined in Section~\ref{sec:binary_lens_parameters}.}
	\label{fig:causticflow_overview}
\end{figure*}

\model combines the advantages of Neural CDEs \citep{kidger2020, Kidger2022} and normalizing flows \citep{rezende2015variational, papamakarios2021normalizing}, which have been applied to microlensing modeling separately \citep{zhang2021romanML, zhao&zhu2022}. 
The architecture is summarized in Figure~\ref{fig:causticflow_overview}.
Below we provide a brief description of the different components.

\subsection{Irregular time series with Neural CDEs}
Neural CDEs are controlled differential equations with neural vector fields.
The model utilizes a neural network to approximate the mapping between the observed data and the evolving physical process/hidden states $z_{t_N}$.
Specifically, the evolution of the hidden state is governed by the following Riemann-Stieltjes integral
\begin{equation}
		z_{t_N} = z_{t_1} + \int_{t_1}^{t_N} f_\phi(z_s)\,dX_s,
\end{equation}
where $z_{t_N}$ is the hidden state at the final observed epoch $t_N$, modified in response to the observations, $f_\phi$ is the learned neural network with parameters $\phi$, and $X_s$ is the continuous control path constructed via the Hermite cubic splines interpolation \citep{Morrill2021Hermite_cubic_splines} of the discrete observed data.
Because Neural CDEs rely on the integration of the continuous data representation rather than the discrete data points, the summarized hidden state $z_{t_N}$ is inherently sampling-invariant.
It naturally accommodates irregular sampling, data gaps, and different observational cadences.
This inductive bias effectively generalizes Neural CDE for irregular time series, making it well-suited for microlensing event with large data gaps \citep{zhao&zhu2022}.

Directly integrating the differential equation over raw binary lensing data is computationally prohibitive due to the long data length and stiffness near caustic crossing.
To accelerate the training process, we use the neural rough differential equations (Neural RDEs, \citealt{Morrill2021neuralRDE} ).
Following the formulation in \cite{Morrill2020signature_method, Morrill2021neuralRDE} and the applications in \citet{zhao&zhu2022}, we apply a log-signature transform within a sliding window.
With reduced sequence length, the sequence is more slowly varying over the full interval than the raw data.
This makes the differential equation better behaved than the original \citep{Morrill2021neuralRDE, Kidger2022}, which leads to the speed-up of the training and inference process.
We also perform the pre-signature scaling to improve the numerical stability \citep{Morrill2020signature_method}.

Regarding the implementation, we follow the Neural RDE formulation of
\citet{Morrill2021neuralRDE} and microlensing application by \citet{zhao&zhu2022}, while adopting different hyperparameters and a JAX-based implementation. We compute log-signatures using \texttt{signax} \footnote{\url{https://github.com/anh-tong/signax}} \citep{kidger2021signatory} with log-signature truncation depth $k=4$ and window size $ W=10$ for 1000 points of simulated data (see Section~\ref{subsec:data_generation}).
The windowed transformation is applied to a two-channel path consisting of normalized observation time and transformed normalized flux. It produces 100 window-level representations, each containing the eight log-signature coefficients computed at truncation depth four, resulting in an array of shape $(100,8)$.
The finite-depth log-signature provides a compressed representation of the path within each window. Smaller windows provide finer temporal localization but produce a longer input sequence, whereas increasing the truncation depth incorporates additional higher-order path information while rapidly increasing the feature dimension and computational cost.
This choice provides a practical compromise between computational efficiency and the retention of local light-curve structure.
We utilize \texttt{diffrax} \citep{Kidger2022} to implement the Neural CDE.
The architecture consists of three core components: an initial network to map the observation to the initial hidden state, a neural vector field $f_\phi$, and a readout network to map the terminal hidden state $z_{t_N}$ of $d_z = 512$ to the condition vector of $d_{\mathrm {cond}}=256$ required by the subsequent normalizing flow.
All three components are parameterized by residual networks (ResNet, \citealt{he2016deep}) comprising three identity blocks \citep{he2016identity}, with a hidden layer of $d_{\mathrm {resnet}} = d_z = 512$ and \texttt{ReLU} activations.
The neural vector field applies a final \texttt{Tanh} activation to constrain outputs, ensuring the stability of the integration.
\subsection{Neural posterior estimation with normalizing flows}

Normalizing flows are a family of generative models with tractable distributions that can efficiently draw samples and evaluate probability density.
They describes the transformations (flows) of the probability density function with a series of invertible mappings (bijections) \citep{rezende2015variational, papamakarios2021normalizing}.
With a highly expressive bijection $f_\psi$ (parameterized by neural network weights $\psi$), normalizing flows can transform a simple base distribution, typically a standard multivariate Gaussian $u \sim \mathcal{N}(0, I)$, to the complex target posterior (e.g., the binary lensing posterior with complex degeneracies, as shown in \citealt{zhang2021romanML}).
The conditional probability density of the surrogate posterior $\hat{p}_\psi(\theta \mid c_\phi(\mathcal{D}))$ is given by the change of variables theorem:

\begin{equation}
	\begin{aligned}
		\hat{p}_\psi(\theta \mid c_\phi(\mathcal{D}))
		&= p_u(f_\psi^{-1}(\theta; c_\phi(\mathcal{D}))) \\
		&\quad \times \left| \det \left( \frac{\partial f_\psi^{-1}(\theta; c_\phi(\mathcal{D}))}{\partial \theta} \right) \right|,
	\end{aligned}
\end{equation}
where $p_u$ is the probability density of the initial base distribution, and the Jacobian determinant accounts for the volume changes induced by the bijection.
To train a surrogate posterior, we should minimize the Kullback–Leibler (KL) divergence from the true posterior $p(\theta \mid \mathcal{D})$ to the surrogate posterior $\hat{p}_{\phi,\psi}(\theta \mid \mathcal{D})$ .
This is mathematically equivalent to minimizing the expected negative log-likelihood (NLL) over the distribution of the training dataset
\begin{equation}
	\mathcal{L}(\phi, \psi) = \mathbb{E}_{\theta \sim p(\theta); \mathcal{D} \sim p(\mathcal{D}|\theta)} \left [ - \ln \hat{p}_{\phi,\psi}(\theta \mid \mathcal{D}) \right].
\end{equation}

In the \model framework, we utilize the Masked Autoregressive Flow (MAF) architecture \citep{kingma2016improved, papamakarios2017masked} implemented in \texttt{flowjax} \citep{ward2023flowjax} to perform density estimation.
To ensure sufficient expressivity for capturing the sharp and multimodal features of microlensing posteriors, we construct the MAF with $10$ transformation layers.
Each layer employs a Rational Quadratic Spline (RQS) \citep{durkan2019neural} as the bijective unit, configured with $32$ knots. This RQS differs from the affine bijection adopted by \citet{zhang2021romanML} and can provide more flexible transformations within each autoregressive layer.
With sufficient MAF expressivity, we adopt a standard Gaussian as the base distribution.

\section{Data generation and training}\label{sec:data_generation_and_training}
\subsection{Binary lensing parameters}\label{sec:binary_lens_parameters}
The standard parameter set describing the binary lens light curve includes nine parameters:
\begin{equation}
	(t_0, t_{\mathrm{E}}, u_0, \rho, q, s, \alpha, F_{\mathrm{base}}, f_{\mathrm{s}}).
\end{equation}
Here the first three parameters $(t_0, t_{\mathrm{E}}, u_0)$ represent the standard point-source point-lens (PSPL, \citealt{paczynski1986}) parameters: \(t_0\), the time of closest lens-source alignment; \(u_0\), the closest distance of the source to the coordinate origin in units of angular Einstein radius \( \theta_\mathrm{E} \); \( t_\mathrm{E} \), the Einstein radius crossing time.
This timescale is related to the Einstein radius \( \theta_\mathrm{E}\) and the total lens mass \(M_\mathrm{L}\) by:
\begin{equation}
	t_{\mathrm{E}} \equiv \frac{\theta_\mathrm{E}}{\mu_\mathrm{rel}}; \qquad
	\theta_\mathrm{E} \equiv \sqrt{\kappa M_\mathrm{L} \pi_\mathrm{rel}}.
\end{equation}
Here \( \kappa \equiv \frac{4G}{c^2 \mathrm{au}} \simeq 8.144\mathrm{mas}/M_\odot \), and \( (\pi_\mathrm{rel}, \mu_\mathrm{rel}) \) are the lens-source relative parallax and proper motion, respectively \citep{Gould2000}.
The parameter \(\rho \) is the angular size of the source, \(\theta_\star\), in units of $\theta_{\mathrm{E}}$.
The next three parameters $(q, s, \alpha)$ characterize the binary geometry: $s$, the projected binary separation in units of $\theta_{\mathrm{E}}$; $q$, the secondary-to-primary mass ratio; and $\alpha$, the angle between source trajectory and binary lens axis.
The last two parameters, $( F_{\mathrm{base}}, f_{\mathrm{s}})$, represent the flux parameters, with $F_{\mathrm{base}}$ the baseline flux and $f_{\mathrm{s}}$ the fraction of source flux in the total baseline flux.

We normalize the flux to align light curves with different blending fractions and baseline fluxes on a common scale: 
\begin{equation}\label{eq:flux}
	F_{\mathrm{norm}}(t) = \frac{F(t)-\operatorname{Median}(F)}{\operatorname{IQR}(F)}. 
\end{equation}
Here $F(t)$ is the observed light curve in flux, and ${\operatorname{IQR}(F)} = Q_{75}(F) - Q_{25}(F)$ is the interquartile range, defined as the difference between the 75th and 25th percentiles of the data.
This robust normalization largely removes the dependence on the flux parameters and thereby improves generalization across different flux configurations.
Although this normalization discards amplitude information, we observe no substantial degradation in the accuracy of the network's $u_0$ proposals toward lower $f_s$ in the simulated test set. If needed, this information could be retained by providing the IQR normalization scale as an auxiliary network input.

Motivated by the asinh magnitude formalism of \citet{lupton1999}, we apply a simple inverse-hyperbolic-sine transformation to the normalized flux in order to compress the dynamic range while retaining finite values for non-positive normalized fluxes.
\begin{equation}\label{eq:asinh_magnitude}
      m_{\mathrm{asinh}}(t)
      =
      -\frac{2.5}{\ln 10}
      \operatorname{asinh}\!\left[F_{\mathrm{norm}}(t)\right].
  \end{equation}
Before passing the light curve to the network, we also use the normalized time coordinate $t_{\mathrm{norm}}=t/100$. These transformations define the network input as the irregular time series $\mathcal{D}$ used by \model.

Regarding the choice of coordinates, we set the origin at the center of magnification (or the primary caustic) rather than the center of mass.
The offset from the center of the primary caustic to the center of mass is given by \citep{Stefano1996, An2002,chung2005}
\begin{equation}
	\Delta x =
	\begin{cases}
		\displaystyle \frac{q}{1+q}\left(s-\frac{1}{s}\right), & \text{if } s > 1 \\
		0,                                                     & \text{if } s < 1
	\end{cases}.
\end{equation}
As argued by \cite{zhang2021romanML,zhao&zhu2022}, there are several advantages of center of magnification coordinates.
Because the $(t_0,u_0)$ parameters match the morphology of the light curve, this will keep the parameters $(t_0,u_0)$ the same for the close/wide degeneracy, thereby aligning the solutions and preventing artificial multimodality in the $(t_0,u_0)$ subspace.
This can also help mitigate correlations in the relevant parameter space.
Both make the posterior easier for the network to learn.

The parameter prediction in \model is not designed to cover the full nine-dimensional binary lens parameter space.
Instead, \model predicts a subset of six parameters, $(t_{\mathrm{E}}, u_0, \rho, q, s, \alpha)$, while treating $(t_0,F_{\mathrm{base}},f_{\mathrm{s}})$ as nuisance parameters.
Since these nuisance parameters are sampled in the training simulations (Section~\ref{sec:data_generation_and_training}) but omitted from the prediction target, the network learns the corresponding marginal posterior over the training prior.
Consequently, the surrogate posterior does not explicitly represent correlations between the predicted and nuisance parameters.
But conditionally estimating the nuisance parameters for each network proposal during downstream fitting allows the resulting joint samples to recover the dominant correlations.

First, we estimate $t_0$ for each proposal using a simplified template-matching procedure in which only the temporal offset is varied.
The matching process consists of three steps.
Given a raw light curve, we first obtain a rough estimate of $\hat{t}_0$ by identifying the midpoint of the sliding window, which corresponds roughly to the epoch with the maximum flux value.
We then combine $\hat{t}_0$ with the other parameters predicted by \model to generate a template light curve on a dense time grid and construct an interpolation function for efficient evaluation at different values of $t_0$.
The exact offset between the initial estimate $\hat{t}_0$ and the true value $t_0$ is then determined by a brute-force search over a range spanning half of the time window.
Because this procedure requires only one light curve computation, it is both fast and robust.
Finally, the two flux parameters are determined by linear regression conditional on the other parameters of each proposal.

This parameterization differs from that of \citet{zhang2021romanML}, who jointly inferred eight parameters including $t_0$ and $f_{\mathrm{s}}$, and from that of \citet{zhao&zhu2022}, who fixed $\rho$ and used separate networks to infer $(t_0,t_{\mathrm{E}})$ and $(u_0,q,s,\alpha,f_{\mathrm{s}})$.
Because temporal shifts governed by $t_0$ do not alter the astrophysical parameters (e.g., $q$ and $s$), template matching removes the need for the network to learn equivariance over a broad range of temporal translations.
Together with flux normalization, this simplification improves prediction accuracy in our experiments.

\subsection{Data generation}\label{subsec:data_generation}

Our target application is to model binary lens events with relatively high mass ratios ($q>10^{-3}$) from a KMTNet-like survey. 
We therefore draw binary lens parameters from broad priors with $q \geq 10^{-3}$, generate light curves under a simple KMTNet-like cadence and noise model, and retain only events that show significant deviations from a single lens fit.

In the simulation, each time series is generated with 1000 randomly sampled points within a 100-day window centered at zero, corresponding to an average cadence of 0.1\,days. 
The simulations use only single-band $I$-band light curves, without multi-band color information.
We adopt a simple KMTNet-like Gaussian noise model to simulate the photometric uncertainties, with the magnitude error given by
\begin{equation}
	\begin{gathered}
		\sigma_m = \sqrt{\sigma_{\mathrm{sys}}^2 + \sigma_{\mathrm{pho}}^2}; \quad
		\sigma_{\mathrm{pho}} = 10^{0.4(m - m_0)}.
	\end{gathered}
\end{equation}
Here $\sigma_{\mathrm{sys}} = 0.006$ mag approximates the systematic error floor, the exponent $0.4$ sets the scaling of the background-dominated noise, and $m_0 = 21.755$ mag is the reference magnitude.
These parameters are chosen to match the $I$-band photometric noise
curve of data obtained with the KMTNet telescope at the South African
Astronomical Observatory (KMTS) using the updated pySIS pipeline \citep{yang2024}.

The binary lens parameters and baseline magnitude are drawn from the following prior distributions. 
In this paper, $\log$ denotes $\log_{10}$:
\begin{equation}\label{eq:prior}
	\begin{gathered}
		t_0 \sim \text{Normal}(0,16.7) \\
		t_{\mathrm{E}} \sim \text{TruncLogNorm}(10^{1.15}, 10^{0.45}; 5, 100)\\
		u_0 \sim \text{Uniform}(0, 2) \\
		q \sim \text{LogUniform}(10^{-3}, 1) \\
		s \sim \text{LogUniform}(10^{-0.7}, 10^{0.7}) \\
		\alpha \sim \text{Uniform}(0^\circ, 360^\circ) \\
		\rho \sim \text{LogUniform}(10^{-4}, 10^{-2}) \\
		f_s \sim \text{Uniform}(0.01, 1)\\
        m_\mathrm{baseline} \sim \text{Uniform}(16,19).
	\end{gathered}
\end{equation}
Here $t_0$ is measured relative to the center of the 100-day prediction window.
The prior on $t_0$ is motivated by the center-of-magnification coordinates, which is different from the uniform distribution in \citet{zhang2021romanML} and \citet{zhao&zhu2022}.
For most events, especially primary caustic crossing or caustic approach events, $t_0$ can be estimated from the midpoint of a sliding window centered on the peak flux (see Section~\ref{sec:binary_lens_parameters}).
The standard deviation is chosen such that $3\sigma$ corresponds to half of the window size, allowing events with caustic features displaced from the window center to remain covered.
The truncated log-normal prior for $t_{\mathrm{E}}$ approximates the observed distribution from OGLE-IV data \citep{mroz2017, mroz2019}.
The mass ratio is limited to $q \geq 10^{-3}$ to focus on relatively massive binaries with strong caustic features.
Under the adopted KMTS noise curve, this baseline-magnitude range corresponds to an approximate baseline signal-to-noise ratio of $12$--$120$.

\begin{figure}[t]
	\centering
	\includegraphics[width=\linewidth]{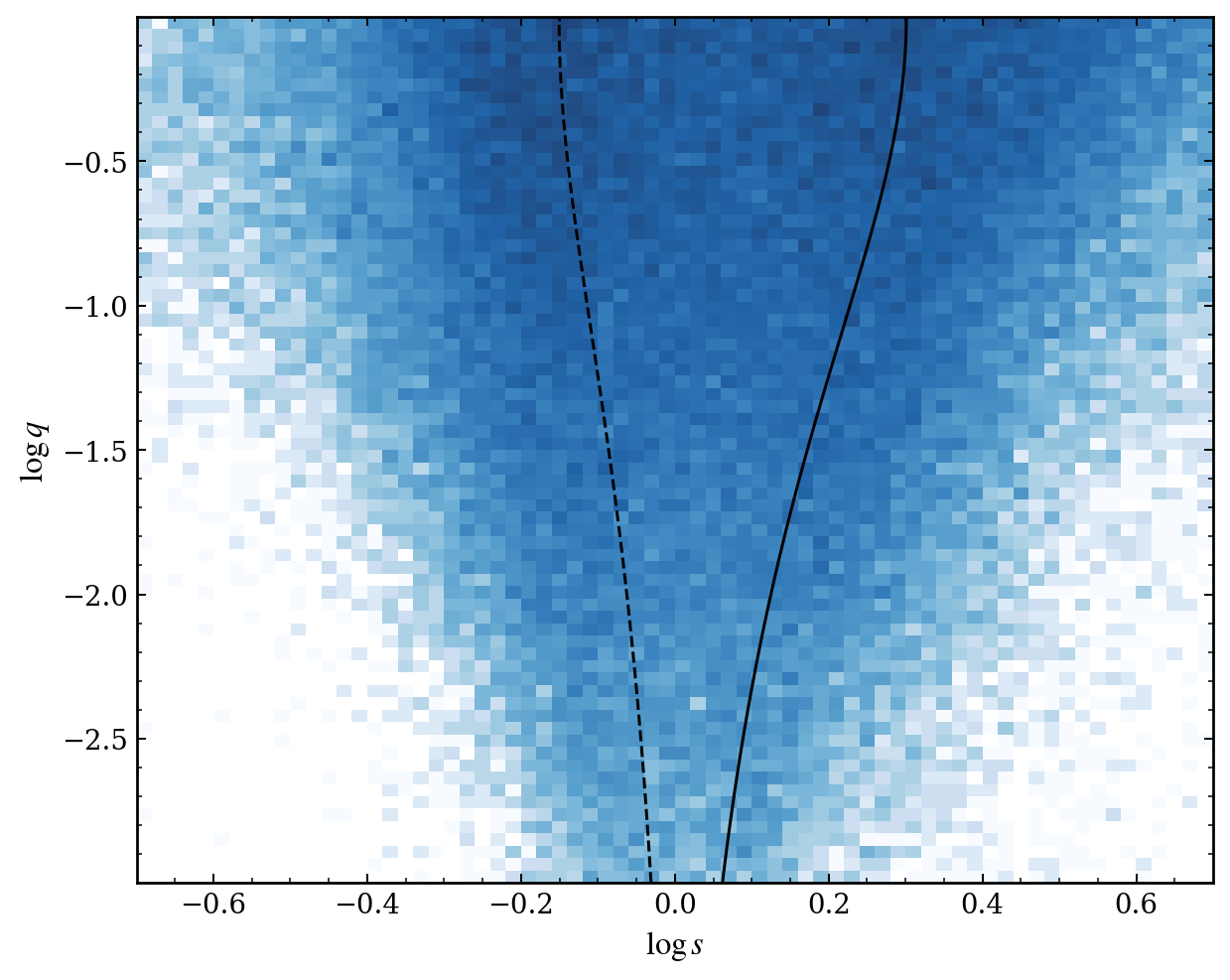}
	\caption{Distribution of simulated binary lens events retained after the single-lens rejection cut, $\chi^2_{\mathrm{PSPL}}/\mathrm{dof} > 2$, in the $(\log s, \log q)$ plane.
		The selection shifts the sample toward parameter regions with stronger binary signatures, while still preserving broad coverage of the close, resonant, and wide caustic topologies.
		The two black curves mark the boundaries between resonant and close/wide caustic geometries \citep{Dominik1999}.
		The retained sample is distributed as $30.37\%$ close, $38.96\%$ resonant, and $30.66\%$ wide.
	}
	\label{fig:topology_distribution}
\end{figure}

We simulate $2\times10^6$ binary lens light curves using \texttt{microlux} \citep{ren2025}, which supports Fisher matrix calculations through automatic differentiation and provides an instant diagnostic comparison between the learned posterior and Fisher-approximate posterior.
For each simulated light curve, we fit a single-lens model and retain only those with reduced chi-square $\chi^2_{\mathrm{PSPL}}/\mathrm{dof} > 2$.
As shown in Figure~\ref{fig:topology_distribution}, this selection ensures that the training set contains strong binary signatures and effectively shifts the parameter distribution toward cases with larger caustics, such as those with larger mass ratios or smaller impact parameters.

\subsection{Training}

We train \model using the AdamW optimizer \citep{Loshchilov2019} in \texttt{optax}, with global gradient clipping \citep{pascanu2013difficulty} to stabilize optimization and weight decay ($10^{-4}$) applied to the linear-layer weights.
The $2\times10^6$ simulated light curves are divided into 20 datasets, each containing $10^5$ samples.
We define one training step as a full pass over a single dataset.
Using a batch size of 1024, the learning rate follows a warm-up cosine decay schedule: it increases linearly from 0 to $5\times10^{-4}$ over the first 5 steps and then decays to a minimum of $5\times10^{-6}$ following a cosine schedule.
Within each dataset, we use a 95/5 train-validation split and apply early stopping when the validation loss does not decrease for 20 steps to prevent overfitting.
A separate dataset of $10^5$ simulated light curves, generated using
the same procedure, is reserved exclusively for testing.
Training stops after 245 steps and requires 28 hours on a single NVIDIA GeForce RTX 5090.
The NLL decreases from 9.68 to $-7.40$ on the training set, while reaching $-6.79$ on the validation set.

\section{Results on Simulated Data}\label{sec:simulation_results}

\begin{figure*}[htb!]
	\centering
	\includegraphics[width=\linewidth]{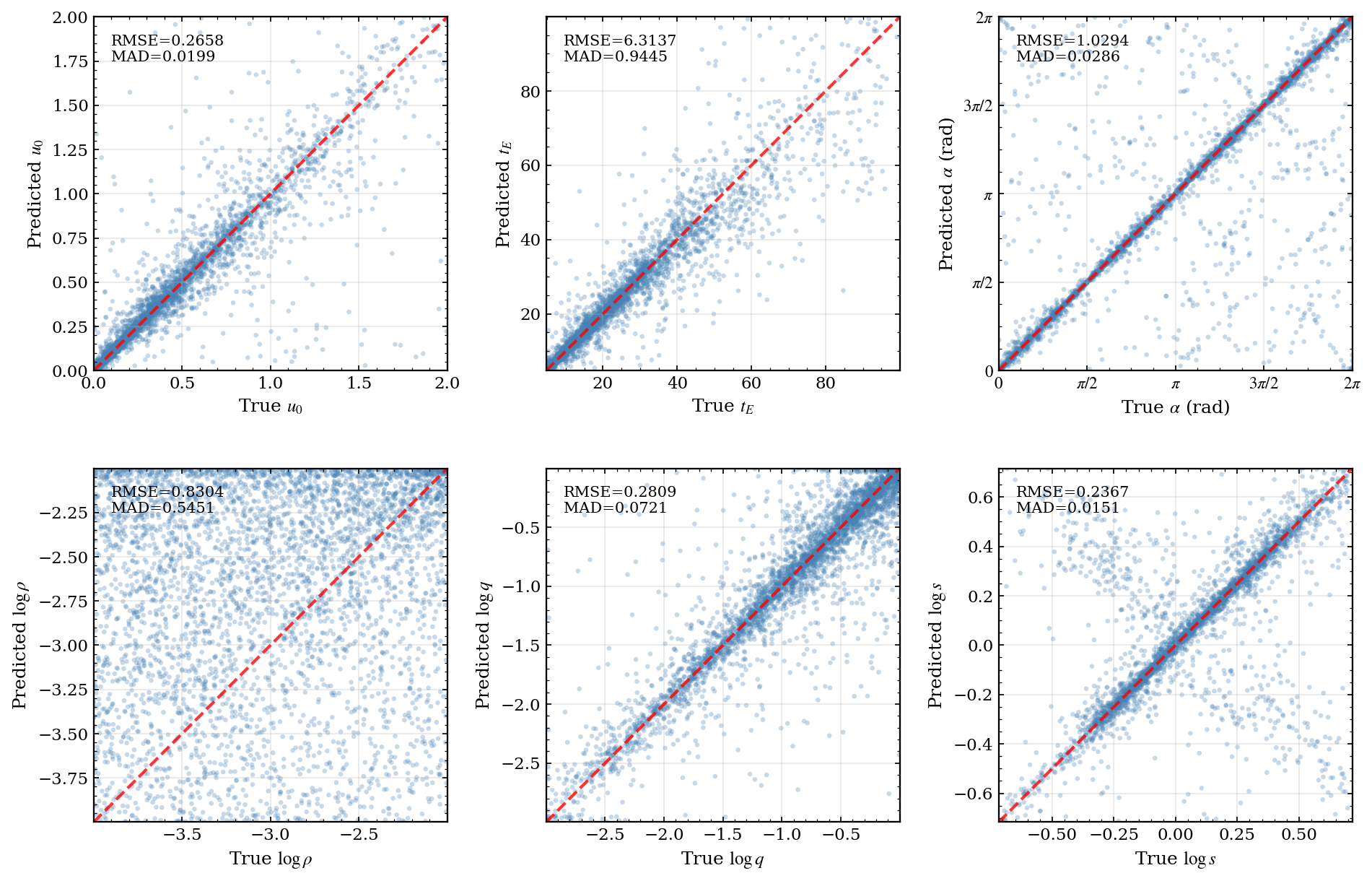}
	\caption{True versus MAP-predicted parameters on the held-out simulated test set for $u_0$, $t_{\mathrm{E}}$, $\alpha$, $\log \rho$, $\log q$, and $\log s$.
		The red dashed lines indicate the one-to-one relation, and the RMSE and MAD are listed in each panel.
		The off-diagonal structures in the $\alpha$ and $\log s$ panels trace the learned physical degeneracies, while the broader dispersion in $\log\rho$ indicates that this parameter is predicted less precisely than the other parameters.
	}
	\label{fig:truth_vs_prediction}
\end{figure*}

The trained model can generate $10^5$ posterior samples in approximately 1 GPU second, making it substantially faster than traditional pipelines based on grid search and MCMC.
This computational advantage, however, must be accompanied by a careful validation of the predictive performance of \model, particularly its ability to generalize to real microlensing events.
In this section, we evaluate the performance of \model on simulated data from several complementary perspectives.

We evaluate the predictive performance of \model on a held-out simulated test set containing $10^5$ light curves.
For each light curve, we normalize the flux using Equation~(\ref{eq:flux}, \ref{eq:asinh_magnitude}) and rescale the time axis as $t_{\mathrm{norm}} = t/100$.
We then compute the sliding log-signature with pre-signature scaling \citep{Morrill2020signature_method} to improve the numerical stability of the Neural CDE.
The resulting log-signature features are passed to \model to obtain the learned posterior.
For each event, we choose from the learned posterior the set of parameters corresponding to the maximum-a-posteriori (MAP) as the prediction, and compare it with the ground truth (i.e., the generating parameters).
As shown in Figure~\ref{fig:truth_vs_prediction}, the predicted values of $u_0$, $t_{\mathrm{E}}$, $\alpha$, and $\log q$ follow closely the ground truth along the one-to-one relation, indicating that the model captures the dominant structure of the posterior over a broad range of parameter space.
That figure also reveals physically meaningful degeneracy patterns learned by the model, including the $\alpha \rightarrow \alpha \pm \pi$ degeneracy associated with central symmetry of resonant caustics and the well-known ``close/wide'' degeneracy \citep{Griest1998, Dominik1999, an2005} in $\log s$. 
For the remaining parameter, $\log\rho$, the model prediction shows substantial scatter around the ground truth. This is partly expected because $\rho$ is primarily constrained by localized finite-source features near caustic crossings or approaches, which are sparsely sampled or absent in many of the simulated light curves at the adopted cadence of roughly 10 points per day. This scatter may also reflect information loss during log-signature preprocessing and subsequent feature embedding.

Similar to \citet{zhao&zhu2022}, we evaluate the agreement between the prediction and the ground truth using both the root mean square error (RMSE) and the median absolute deviation (MAD), with both values listed in each panel of Figure~\ref{fig:truth_vs_prediction}. 
As a reference, we compare the MAP prediction with the joint-pipeline performance of \citet{zhao&zhu2022} shown in their Figure~16. 
Although the comparison is not strictly fair because the training sets, parameter ranges, and predicted parameter spaces differ, our MAP predictions reduce the MAD by factors of about $1.5$--$5.4$ for the shared parameters $(u_0,\alpha,\log q,\log s)$.
Taking MAD as a measure of the prediction accuracy, \model is able to recover the key microlensing parameters, namely $\log{q}$ and $\log{s}$, at a level of $<0.1$ dex, which is sufficient for statistical analysis of large ensembles of binary events.

\begin{figure}[htb!]
    \centering
    \includegraphics[width=\linewidth]{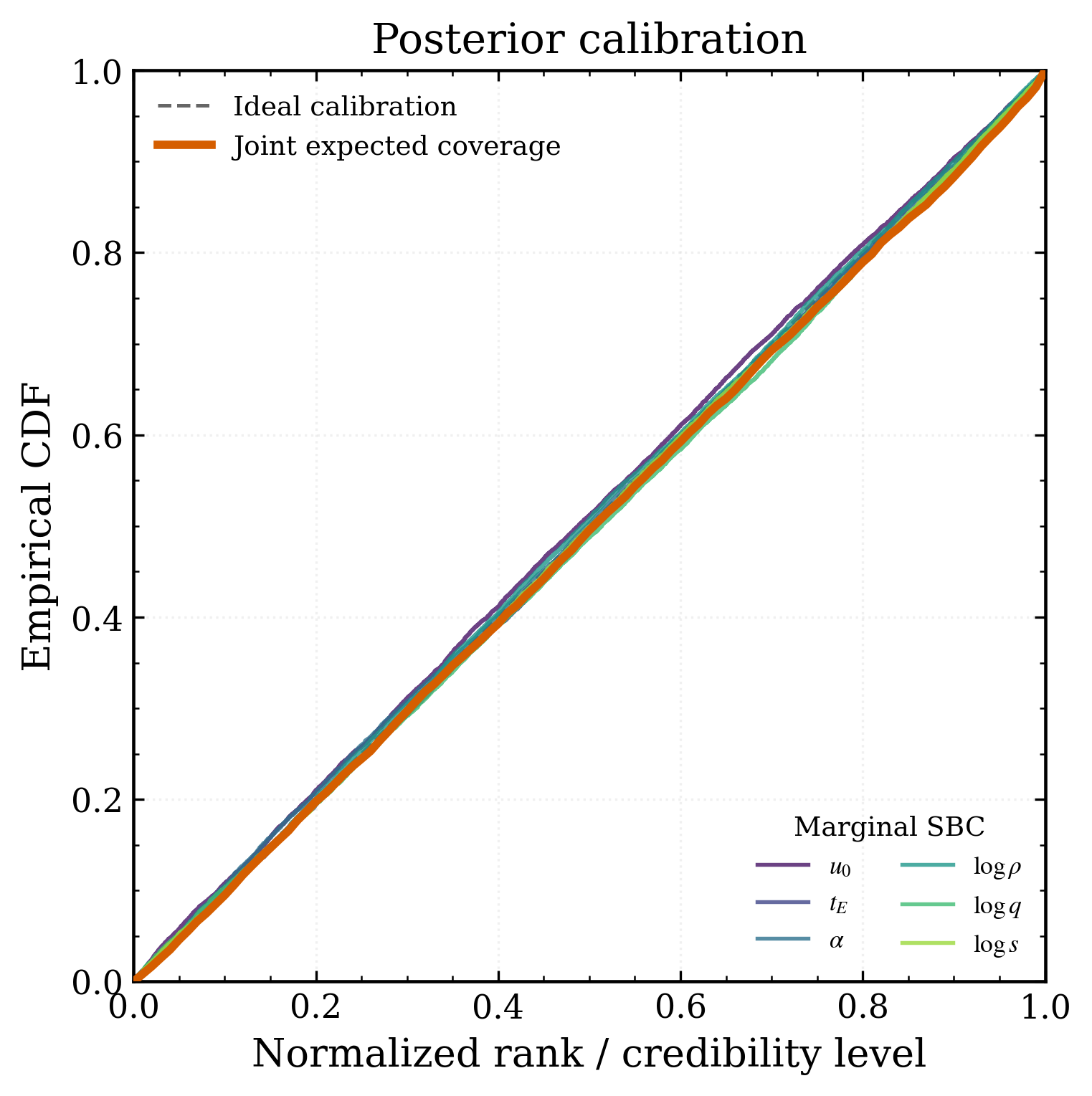}
    \caption{
    Posterior-calibration diagnostics on the held-out simulated test set.
    The colored curves show the empirical cumulative distributions of the normalized SBC ranks for the six inferred parameters, while the orange curve shows the joint expected coverage based on HPD regions.
    The gray dashed line indicates the ideal one-to-one relation for a calibrated posterior.
    }
    \label{fig:posterior_calibration}
\end{figure}

While the MAP estimate summarizes the mode of the posterior, it does not assess whether the posterior uncertainty is statistically calibrated.
We therefore evaluate posterior calibration using both simulation-based calibration (SBC) ranks \citep{talts2018validating} for the marginal posterior of each parameter and joint expected coverage based on highest-posterior-density (HPD) regions \citep{deistler2022truncated}.
For a calibrated posterior, the normalized rank of the true value among posterior samples should be uniformly distributed for each parameter.
Similarly, for the joint posterior, the posterior mass of the HPD region whose boundary passes through the true parameter vector should be uniformly distributed across simulated events. Consequently, the empirical cumulative distribution of both statistics should follow the one-to-one relation.
As shown in Figure~\ref{fig:posterior_calibration}, both the marginal SBC curves for all six parameters and the joint expected-coverage curve closely follow the ideal one-to-one relation, indicating that the posterior uncertainties are well calibrated on average over the selected simulated test population, both marginally and jointly.
In particular, although $\log\rho$ is predicted less precisely than the other parameters, its marginal SBC ranks remain close to the ideal relation, indicating that the corresponding posterior uncertainties are nevertheless well calibrated on average over the simulated test population.

\begin{figure*}[htb!]
	\centering
	\includegraphics[width=\linewidth]{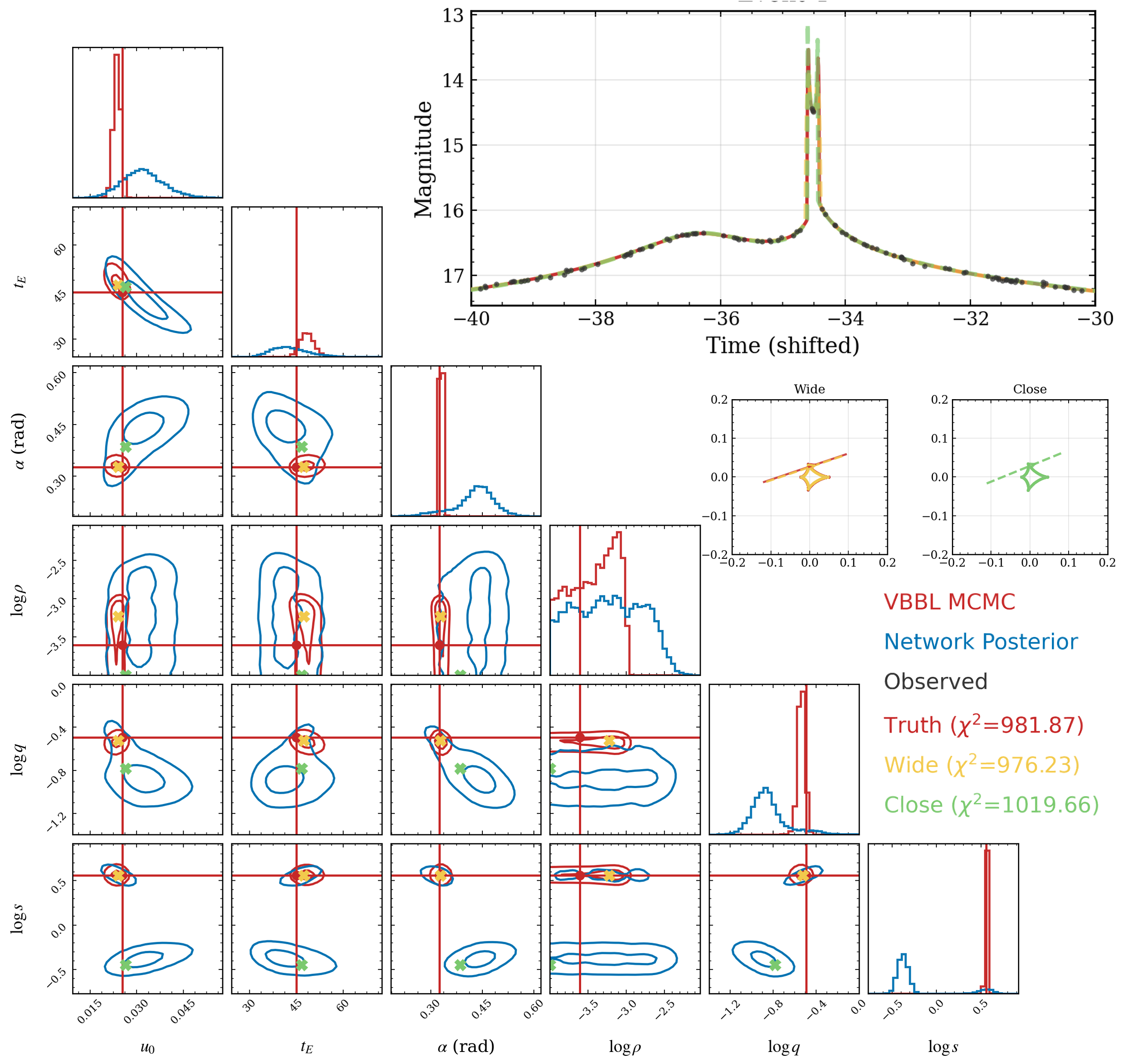}
	\caption{Posterior comparison for a representative multimodal event exhibiting the close/wide degeneracy from the simulated dataset.
		The left corner plot compares the posterior structure inferred from the MCMC (red) with the surrogate posterior from \model (blue) in the six-dimensional parameter subspace $(u_0, t_{\mathrm{E}}, \alpha, \log \rho, \log q, \log s)$.
        The red point and lines mark the true input parameters, while the yellow and green crosses mark the polished wide and close solutions, respectively.
        The upper-right panel compares the corresponding light-curve models with the simulated observations, and the middle-right panels show the source trajectories and caustic geometries.
		The surrogate posterior does not exactly match the local MCMC posterior, but it preserves the two physically relevant solution branches.
		This example illustrates that the surrogate posterior preserves the physically relevant multimodality needed for downstream optimization and interpretation.
	}
	\label{fig:corner_mcmc_vs_network_event_4}
\end{figure*}

These global diagnostics provide necessary conditions for posterior calibration on average over the simulated population. To investigate the local posterior correctness on individual events, we compare the learned posterior with reference posteriors obtained from MCMC.
Figure~\ref{fig:corner_mcmc_vs_network_event_4} shows a representative simulated event exhibiting the close/wide degeneracy.
For this comparison, the MCMC chains are initialized near the true parameters, and the polished solutions are obtained using the posterior-guided optimization procedure described below.
Although the learned posterior does not exactly reproduce the local MCMC posterior, it preserves the physically relevant close and wide solution branches.
Drawing initial points from this posterior and polishing them with local optimization recovers both branches with $\Delta \chi^2<40$ in this example, as shown by the corresponding light-curve fits and lens geometries.

\begin{figure*}[htb!]
	\centering
	\includegraphics[width=\linewidth]{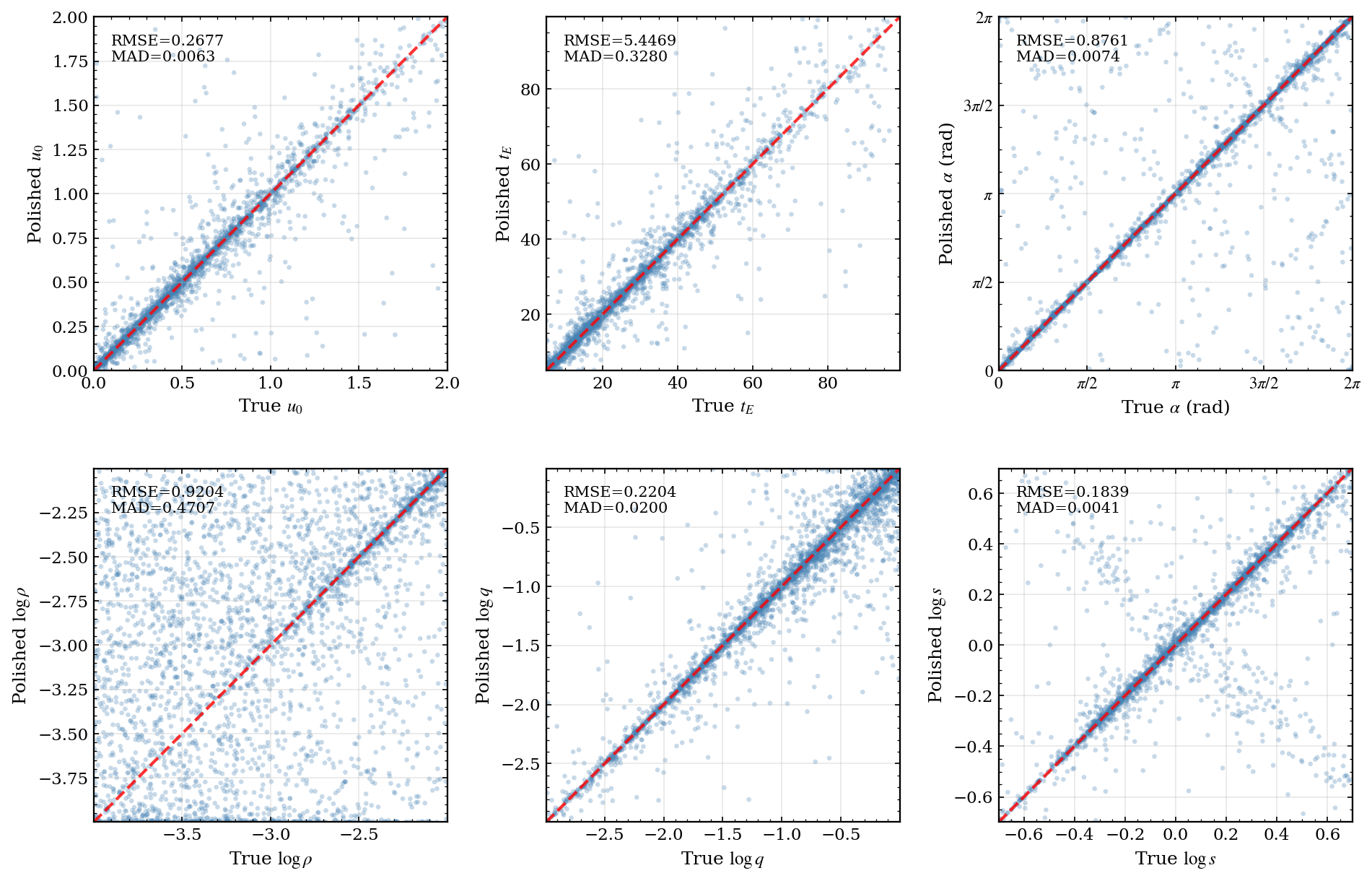}
	\caption{True versus polished parameters on the held-out simulated test set. For each event, $N=10$ initial points are drawn from the learned posterior and refined by local optimization, and the best-polished solution is shown. The layout is the same as in Figure~\ref{fig:truth_vs_prediction}.
	}
	\label{fig:polished_vs_truth}
\end{figure*}

\begin{figure}[htb!]
	\centering
	\includegraphics[width=\linewidth]{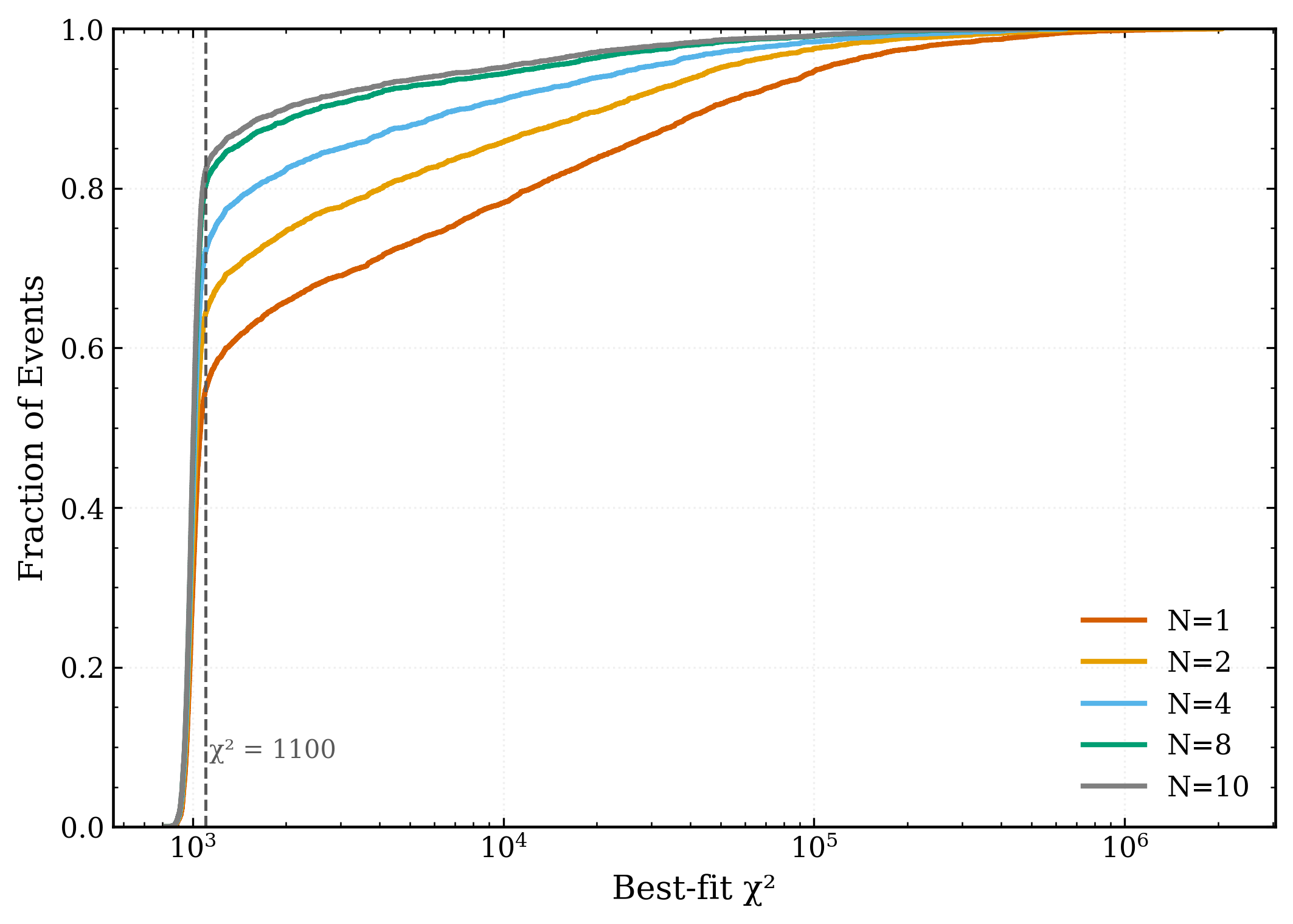}
	\caption{Cumulative distributions of the best-fit $\chi^2$ values obtained after local minimization from $N$ initial points drawn from the learned posterior.
		Different curves correspond to different numbers of starting points, up to $N=10$. The vertical dashed line marks $\chi^2 = 1100$.
	}
	\label{fig:chi2_cumulative_by_n}
\end{figure}

The above example also underscores the potential of using the learned posterior of \model as an informative proposal distribution for global parameter optimization, rather than a replacement for full MCMC or Hamiltonian Monte Carlo modeling in individual events. 
Specifically, for a given light curve, \model yields a learned posterior from which we draw $N$ initial points.
Each of these points is then refined with a local optimizer such as the Nelder-Mead simplex algorithm \citep{nelderSimplexMethodFunction1965, gaoImplementingNelderMeadSimplex2012} and nonlinear least-squares method \citep{branch1999subspace}, with \texttt{VBMicrolensing} as a backend model \citep{bozza2010,bozza2025} for its computational efficiency. This workflow exploits the multimodality and parameter correlations learned by the network.
Figure~\ref{fig:polished_vs_truth} compares the true parameters with the best-polished solutions obtained from $N=10$ posterior-drawn initial points. 
Compared with the MAP predictions in Figure~\ref{fig:truth_vs_prediction}, posterior-guided polishing substantially improves the agreement for the well-constrained parameters, with an improvement in MAD by a factor of $\sim3$ in both $\log{q}$ and $\log{s}$.
These optimizations also improve the goodness of fit of the model. As shown in Figure~\ref{fig:chi2_cumulative_by_n}, the recovery fraction---defined by $\Delta \chi^2<100$ between the best-fit and input models---increases with the number of posterior-drawn starting points, reaching about $80\%$ with the adopted budget of ten local optimizations on simulated data.
For comparison, a traditional grid search typically requires more than $10^4$ trial evaluations.
This suggests that our modeling procedure can achieve a substantial speed-up during the optimization stage.


\section{Application to Real Events}\label{sec:real_events}
We verify the performance of \model on a set of 10 real microlensing events observed during the 2023--2025 seasons and reported by \citet{Han2026}. These events are chosen because the binary lenses have mass ratio $\log{q}>-3$, thus matching the mass-ratio range used for our training samples.
That said, the chosen events show deviations from the simulated events in several other aspects, as listed in Table~\ref{tab:real_event_summary} and detailed below:
\begin{itemize}
    \item Event timescale $t_{\mathrm{E}}$: Both KMT-2025-BLG-0922 ($t_{\mathrm{E}}=2.68\,$days) and KMT-2025-BLG-1056 ($t_{\mathrm{E}}=4.58\,$days) have shorter timescales than the shortest timescale (5\,days) of simulated events.
    \item Source radius $\rho$: KMT-2025-BLG-0922, KMT-2025-BLG-1056, and KMT-2025-BLG-2427 all have $\rho > 10^{-2}$, beyond the upper limit of $\rho$ used for simulated events.
    \item Higher-order effects: OGLE-2023-BLG-0249 shows both annual parallax effects \citep{Gould1992} and lens orbital motion effects \citep{dominik1997galactic,albrow2000detection,skowron2011binary}, and KMT-2025-BLG-2427 shows strong lens orbital motion effect. Neither of these effects was included in the simulated events.
    \item Sampling cadence and photometric behavior: The sampling cadences of these events range from $0.026$ to $0.48$ days, with a mean cadence of $0.061$ days, whereas the simulated training light curves have an average cadence of $0.1$ days. In addition, the training samples are generated under the assumption of Gaussian white noise, while real observations can contain correlated noise and other photometric systematics.
\end{itemize}
Given all these differences, we did not expect \model to match its performance on simulated events. Nevertheless, as shown below, it still recovers most of the real events in our test sample. 
This test therefore demonstrates the generalizability of \model, in addition to helping identify directions for future improvement (see Section~\ref{sec:discussion}).

The real light-curve data are preprocessed before being passed to \model. We start from the aligned light curves from C.\ Han (private comm.); this light curve alignment is in general independent of the microlensing model, although for this test we have made use of the best-fit models by \citet{Han2026} for simplicity.
We first remove high-uncertainty measurements using a threshold of three Gaussian-normalized MADs above the median photometric uncertainty.
To suppress isolated photometric outliers without removing possible anomaly points, we identify and protect prominent brightening features before applying local outlier rejection.
We first smooth the magnitude light curve with a Savitzky--Golay filter \citep{savitzky1964smoothing} in \texttt{scipy.signal.savgol\_filter} using a window length of 51 data points and a polynomial order of three.
We then identify peaks in the negative smoothed magnitude using \texttt{scipy.signal.find\_peaks}, with a minimum prominence of 2.5 times its Gaussian-normalized MAD and a minimum width of one data point.
We retain peaks with prominences of at least 20\% of that of the strongest peak and define protected windows from their half-prominence widths, with an additional 2-day buffer on each side.
Finally, outside these protected windows, we apply a Hampel filter \citep{hampel1974influence} with a window size of 10 data points and a $3\sigma$ threshold.
After this cleaning step, we identify the input segment for network prediction by searching for the 100-day window with the largest integrated flux excess relative to the baseline.
The selected light curve segment is then normalized using the procedure described in Section~\ref{sec:simulation_results} and passed to \model to obtain the learned posterior.
Because of the model mismatch, we choose to draw 20 random samples from the learned posterior for the polishing stage, which is more than in the simulated-data test.
These samples are then refined against the full light-curve data using the nonlinear least-squares method of \citet{branch1999subspace}. The parameter uncertainties are then estimated via the Fisher matrix method, making use of the model derivatives of \texttt{microlux} \citep{ren2025}.
The full workflow takes about 1 GPU second, or 5 CPU seconds, per event for network prediction, followed by about 10 CPU minutes per event for the subsequent polishing.

\begin{deluxetable*}{lrrllll}
    \tabletypesize{\small}
    \tablecaption{Summary of real-event fits.\label{tab:real_event_summary}}
    \tablewidth{0pt}
    \tablehead{
        \colhead{Event} &
        \colhead{$\Delta\chi^2/\mathrm{dof}$} &
        \colhead{$I_{\rm base}$} &
        \colhead{Degeneracy} &
        \colhead{Higher-order} &
        \colhead{Outside training range} &
        \colhead{Outcome}
    }
    \startdata
    OGLE-2023-BLG-0249 & 0.0/840 & 15.69 & \nodata & parallax, orbital & \nodata & recovered \\
    KMT-2023-BLG-1246 & 313.4/1508 & 19.82 & inner/outer & \nodata & \nodata & failed \\
    OGLE-2023-BLG-0079 & 0.0/2824 & 19.50 & inner/outer & \nodata & \nodata & recovered \\
    KMT-2024-BLG-0072 & 0.0/271 & 19.77 & \nodata & \nodata & \nodata & recovered \\
    KMT-2024-BLG-0897 & 11.9/5459 & 18.16 & \nodata & \nodata & \nodata & recovered \\
    KMT-2024-BLG-1876 & -7.1/9164 & 18.89 & \nodata & \nodata & \nodata & recovered \\
    KMT-2024-BLG-2379 & 45.4/3218 & 19.89 & close/wide & \nodata & \nodata & recovered \\
    KMT-2025-BLG-0922 & -0.6/2233 & 17.45 & \nodata & \nodata & $t_{\rm E}=2.68$; $\log\rho=-1.22$ & recovered \\
    KMT-2025-BLG-1056 & 2922.0/2949 & 19.18 & \nodata & \nodata & $t_{\rm E}=4.58$; $\log\rho=-1.37$ & failed \\
    KMT-2025-BLG-2427 & 289.9/527 & 18.91 & \nodata & orbital & $\log\rho=-1.752$ & failed \\
    \enddata
    \tablecomments{$\Delta\chi^2=\chi^2_{\mathrm{best\ polished}}-\chi^2_{\mathrm{reference}}$.
        The outcome is determined by the combined criterion of $\Delta\chi^2<50$ and close agreement with the reference lensing geometry.
        The “Outside training range” column indicates parameters that fall beyond the prior range used in the training simulations.}
\end{deluxetable*}

\begin{figure*}[htb!]
	\centering
	\includegraphics[width=\linewidth]{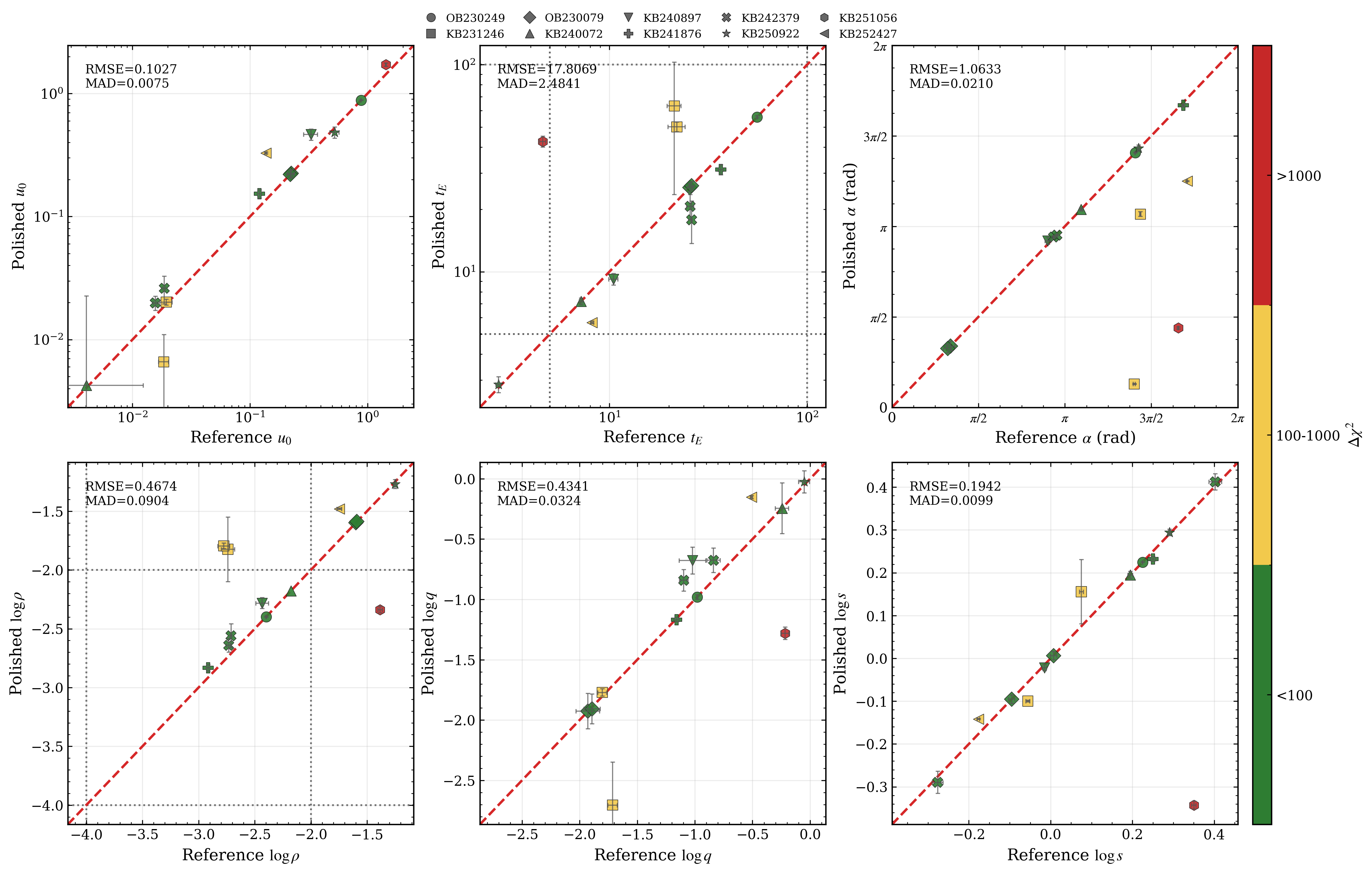}
	\caption{Comparison between the reference parameters and the corresponding polished parameters obtained from the learned posterior for the real-event test sample.
	The reference parameters are obtained by transforming the literature solutions into our parameter convention and refining them with a static binary lens model. Hence they are not the same as given in the discovery paper \citep{Han2026}.
	The red dashed lines mark the one-to-one relation, and different symbols denote different events. For each of the three events with a discrete degeneracy, each matched reference--polished solution pair is plotted as a separate point (Table~\ref{tab:real_event_summary}).
	Solutions are color-coded by the goodness-of-fit difference between the polished and the reference models, $\Delta\chi^2=\chi^2_{\mathrm{polished}}-\chi^2_{\mathrm{reference}}$.
    Gray error bars indicate the parameter uncertainties: horizontal bars show the transformed literature uncertainties from \citet{Han2026}, and vertical bars show approximate uncertainties of the polished solutions estimated from the local Fisher information matrix with \texttt{microlux} \citep{ren2025}.
    For events without constrained finite-source effects in the literature solutions, the plotted $\log\rho$ values are formal values returned by the static-model minimization, and the corresponding $\log\rho$ error bars are omitted.
    The vertical and horizontal gray dotted lines mark the boundaries of the corresponding parameters used in the simulation.
	The RMSE and MAD of the parameter comparisons are given in each panel.
	}
	\label{fig:real_event_parameter_comparison}
\end{figure*}

To provide a fair comparison, we transform the literature solutions of \citet{Han2026} into our parameter convention, using the center-of-magnification coordinate system and the equivalent branch with $u_0>0$ and $\log q<0$. We then refine them with a static binary lens model (i.e., no lens orbital motion or annual parallax effects). These serve as the reference solutions.
We regard an event as recovered if the polished solution has a $\chi^2$ value close to the reference solution, defined as $\Delta \chi^2<50$ and the lensing geometry matches closely that of the reference solution.
Under these criteria, 7 of the 10 events are recovered (see Table~\ref{tab:real_event_summary}).
Figure~\ref{fig:real_event_parameter_comparison} compares the recovered parameters in our polished solutions with the reference values.
The recovered events generally lie close to the one-to-one relation, whereas the failed cases show larger deviations in at least some of the model parameters.
For the seven recovered events, the Fisher-matrix uncertainties of the polished solutions are consistent with the transformed literature uncertainties to within a median factor of $1.7$, supporting their use as approximate uncertainty estimates.

\begin{figure*}[htb!]
	\centering
	\includegraphics[width=\linewidth]{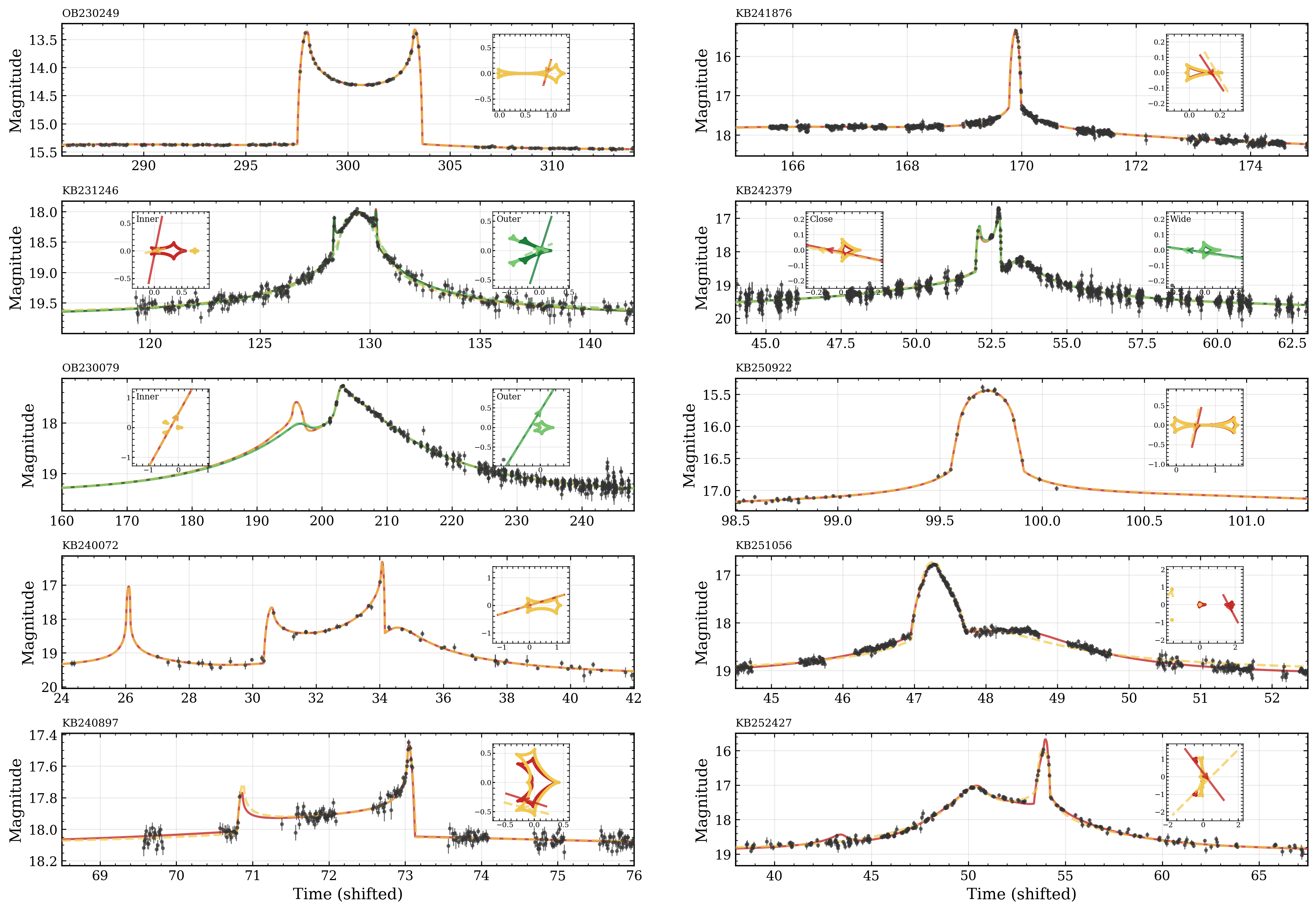}
	\caption{Light-curve and lens-geometry comparison for the real-event test sample.
    Black points with error bars show the observations. The red and green lines denote the reference solutions for the first and second degenerate branches, respectively, while the yellow and light-green lines denote their corresponding polished solutions. The insets show the corresponding source trajectories relative to the caustic structures using the same color scheme.
	}
	\label{fig:real_event_atlas}
\end{figure*}

We show in Figure~\ref{fig:real_event_atlas} the light curves and lensing geometries of the 10 real events. More details of the light curves as well as event modeling are given by \citet{Han2026}. Below we discuss the 10 events individually.
\begin{itemize}
    \item OGLE-2023-BLG-0249. 
    This event shows a typical U-shaped light curve and has the best data quality in the test sample.
    It nevertheless shows prominent higher-order effects including both parallax and orbital motion, which improve the detailed model $\chi^2$ by $\sim370$ \citep{Han2026}. 
    Despite this mismatch between the data and our simulation, our pipeline recovers this event robustly. In fact, no further polishing is needed in this case, as \model alone is able to predict the microlensing parameters fairly accurately.

    \item KMT-2023-BLG-1246. 
	This is a challenging case because its full light curve can be approximated by a PSPL model, with relatively weak caustic-crossing features.
    The modeling by \citet{Han2026} also revealed an inner/outer degeneracy with no significant $\chi^2$ difference. 
    For this event, the learned posterior from \model is biased away from the reference solutions, most clearly in the parameter $\alpha$ (see Figure~\ref{fig:real_event_parameter_comparison}).
    Because of this bias, the subsequent sampling and local refinement are unlikely to recover the correct solution.
    This failure is plausibly related to the weakness of the binary lens anomaly and the more complicated noise behavior of the real data.

    \item OGLE-2023-BLG-0079.
    This event was only partially covered by ground-based observations as the peak was before the start of the microlensing season, making it a useful test of the robustness of \model in dealing with large data gaps (see also \citealt{zhao&zhu2022}).
    According to \citet{Han2026}, this event exhibits an inner/outer degeneracy with $|\log{s}|\approx0.1$, with neither solution showing prominent finite-source effects.
    Because the two degenerate solutions are close in the parameter space, the learned posterior does not separate the inner and outer modes into a clear bimodal structure. Nevertheless, the subsequent polishing is able to identify and successfully recover the two degenerate solutions.
    According to our modeling, the two degenerate solutions show clear differences around the shifted time of $\sim195$\,days, but this part of the light curve is not observed because of the missing data (see Figure~\ref{fig:real_event_atlas}).
    
    As an additional proof-of-concept test, we reversed the time axis of this light curve, thereby turning its observed latter portion into the leading portion of a partially observed event. The network-guided workflow recovered the expected time-reversed outer solution, suggesting that \model may also be useful for inference on ongoing events.
    However, because the present model was trained and validated primarily on completed events, a robust real-time implementation would require dedicated training on partially observed light curves and causal preprocessing that uses only the observations available at each epoch.

    \item KMT-2024-BLG-0072, KMT-2024-BLG-0897, KMT-2024-BLG-1876, and KMT-2024-BLG-2379.
    These four events are successfully recovered by the network-guided polishing pipeline, with small model differences of $\Delta\chi^2=0.0$, $11.9$, $-7.1$, and $45.4$, respectively. For KMT-2024-BLG-1876, the recovered solution attains an even lower model $\chi^2$, although the small difference of $\Delta\chi^2$ can be attributed to stochastic variation in the polishing stage. Similarly, for KMT-2024-BLG-2379, the recovered solutions are worse than the reference solutions by $\Delta\chi^2=45.4$, which is somewhat large from a statistical point of view. Nevertheless, the microlensing parameters are retrieved reasonably well (Table~\ref{tab:real_event_summary}), and the lensing geometry of the recovered solutions matches the reference solutions closely (Figure~\ref{fig:real_event_atlas}), so we consider it successful. According to \citet{Han2026}, this event also shows close/wide degeneracy, and both solutions are recovered.

    \item KMT-2025-BLG-0922.
    The lens is a nearly equal-mass binary with $q\approx1.1$ \citep{Han2026}, or $q\approx 0.9$ under our definition.
    This results in an almost mirror-symmetric caustic topology, which very often would lead to a geometric degeneracy in $\alpha$.
	Additionally, both $t_{\rm E}$ and $\log\rho$ are outside the parameter ranges used in the simulation. Nevertheless, our pipeline recovers all six parameters as well as the lensing geometry successfully.

    \item KMT-2025-BLG-1056.
    The network-guided polishing pipeline fails catastrophically ($\Delta\chi^2=2922.0$) for this event, with almost all predicted parameters showing substantial deviations from the reference values (Figure~\ref{fig:real_event_parameter_comparison}). The lensing geometry does not match either. As the predicted light curve appears qualitatively similar to the reference one, it is likely that the failure is related to the fact that both $t_{\rm E}$ and especially $\log\rho$ are beyond the corresponding ranges used for simulated events.

    \item KMT-2025-BLG-2427.
	This is a failed case with modest $\Delta\chi^2<300$ (Table~\ref{tab:real_event_summary}). The reference $\log\rho$ lies outside the parameter range of our simulation. More importantly, this event exhibits strong lens orbital motion, which not only improves the model $\chi^2$ by $>200$ but also modifies the light-curve morphology around the shifted time of $43$~days (Figure~\ref{fig:real_event_atlas}, see also \citealt{Han2026}). As the predicted light curve matches the data qualitatively well except for a few data points around the caustic crossing region, the mismatch between the simulation and real event---especially the strong orbital motion effect---is probably the cause of the failure.
\end{itemize}

To summarize, our experiment with the 10 real events shows that \model is able to recover the model parameters for $\sim70\%$ of the events after simple polishing, and that the accuracy (measured by MAD) is at levels of $\sim7\%$ and $\sim2\%$ for mass ratio $q$ and projected separation $s$, respectively. These values match closely those found on the simulated events (Section~\ref{sec:simulation_results}), even though our simulation was not designed to match very closely the properties of the real events. This experiment demonstrates that \model can accommodate several forms of mismatch between simulated and real events, although not all.

\section{Discussion} \label{sec:discussion}

In this work, we present \model, a neural posterior estimation framework that combines Neural CDEs with normalizing flows to handle irregularly sampled time series and complex posteriors in binary microlensing modeling.
Once trained, \model can generate posterior samples in seconds.
On simulated data, the MAP estimates recover the main structure of the posterior distribution over a broad region of parameter space and achieve a typical precision of $\sim17\%$ for the binary mass ratio $q$ and $\sim3\%$ for the projected separation $s$. This level of precision is sufficient for statistical analysis of large datasets.
A comparison with the reference posterior from the MCMC further indicates that the learned posterior from \model preserves important parameter correlations and physically relevant degeneracies, although noticeable differences remain between the learned and reference posteriors in individual events.

The learned posterior is also useful for downstream optimization. Rather than replacing the detailed modeling, \model provides a compact learned posterior from which a small number of initial points can be drawn and refined with conventional minimization methods. On simulated data, this posterior-guided optimization substantially improves the accuracy of the model parameters, reaching levels of $<5\%$ and $<1\%$ in $q$ and $s$, respectively, with only 10 posterior samples. The model $\chi^2$ and light-curve morphology also match the ground truth for $\sim80\%$ of events.

For any machine learning model trained on simulated data, the key challenge lies in its generalization performance to real observations. In this respect, \model performs better than expected and substantially improves upon existing models \citep{zhang2021romanML, zhao&zhu2022}.
When applied to 10 binary microlensing events from a systematic analysis of \citet{Han2026}, \model successfully recovers the model parameters, light-curve morphology, and lensing geometry of seven events, with a computational cost of about 10 CPU minutes per event.
The recovery is achieved despite several mismatches in event properties between the simulated and the real ones. 
The recovered parameters also reach similar levels of precision and accuracy to those found for the simulated cases.
Our work demonstrates the feasibility of using \model for systematic modeling of large microlensing datasets, including those from archival and ongoing ground-based surveys and new data from upcoming space-based surveys, such as Roman \citep{spergel2015wide,penny2019}, China Space Station Survey Telescope \citep{CSST2026}, and ET \citep{Ge2022}.
For such a survey-scale application, an upstream anomaly-detection algorithm, such as the KMTNet AnomalyFinder \citep{Zang2021} or other machine learning based methods, would first be needed to identify candidate binary-lens events for subsequent inference with \model.

Although \model is already suitable for real event modeling, several aspects of the framework could be improved.
First, although the learned posterior is well calibrated on average and preserves the main physical degeneracies, it does not exactly reproduce the local MCMC posterior in the representative comparison, and the prediction of $\log\rho$ remains less precise than those of the other parameters. This indicates that the fidelity and precision of the learned posterior can still be improved. Possible directions include shorter log-signature windows and modifications to the current neural architecture, although their benefits and computational costs require systematic evaluation.
Second, the mismatch between the simulated training data and real observations should be reduced.
Specifically, our current training data uses a simple white noise model with an empirical scaling rule derived from the KMTNet survey.
This simplified noise model does not capture outliers or time-correlated photometric errors, or other survey-specific systematics in the real data, which can reduce the capability of \model in recovering relatively weak binary signals.
The lower average cadence adopted in the simulations relative to the real-event sample may also contribute to the relatively weak constraints on $\rho$.
One way to build a more realistic training set is to inject simulated microlensing signals into real baseline data. By doing so, one ensures realistic noise and cadence distributions.
In addition, \model currently does not include the photometric uncertainty column as a model input, because these uncertainties carry limited information under the simplified noise model.
This choice should be revisited, when training on more realistic data, because flux uncertainties may help distinguish outliers from genuine microlensing signals.
Another source of model mismatch is the absence of higher-order effects in the model training.
Although our tests on the real events show that \model is able to make reasonably good predictions for some events that show higher-order effects, the model fails when the higher-order effects become so prominent that it modifies the light-curve morphology (e.g., KMT-2025-BLG-2427). Therefore, incorporating such higher-order effects into the model training procedure is expected to enhance predictive performance, although this must be carefully balanced against the increased computational and statistical costs associated with exploring a substantially larger parameter space.
Last but not least, our simulation so far focuses on the high mass-ratio end of the binary events, namely $q>10^{-3}$, as such events are far more numerous but far less explored in the current analysis. It may nevertheless be useful to extend \model into the planetary event regime, which remains the primary focus of microlensing searches from both ground-based and space-based surveys.

\begin{acknowledgments}
	The authors thank Cheongho Han for providing the real-event data. We also thank Qingru Hu and Hongjing Yang for assistance with early tests and relevant discussion, and Andy Gould and the anonymous reviewer for comments on the manuscript.
	This work is supported by the National Natural Science Foundation of China (grant No.\ 12133005).
	The authors acknowledge the Tsinghua Astrophysics High-Performance Computing platform at Tsinghua University for providing computational and data storage resources that have contributed to the research results reported within this paper.
    This research has made use of the KMTNet system operated by the Korea Astronomy and Space Science Institute (KASI) at three host sites of CTIO in Chile, SAAO in South Africa, and SSO in Australia. Data transfer from the host site to KASI was supported by the Korea Research Environment Open NETwork (KREONET).
    The authors acknowledge the use of large language models (ChatGPT and GLM) for language polishing. The authors are solely responsible for the scientific content of this manuscript.
\end{acknowledgments}

\software{NumPy \citep{numpy2011,numpy2020}, SciPy \citep{scipy2020}, Matplotlib \citep{matplotlib2007,matplotlib_2016}, JAX \citep{jax2018github}, FlowJAX \citep{ward2023flowjax}, Optax \citep{deepmind2020jax}, Equinox \citep{kidger2021equinox}, Diffrax \citep{Kidger2022}, signax (\url{https://github.com/anh-tong/signax}), VBMicrolensing \citep{bozza2025}, microlux \citep{ren2025}, emcee \citep{foreman-mackey2013}, corner \citep{corner}, Jupyter \citep{jupyter}}

\bibliography{sample701}{}

\begin{thebibliography}{}
\expandafter\ifx\csname natexlab\endcsname\relax\def\natexlab#1{#1}\fi
\providecommand{\url}[1]{\href{#1}{#1}}
\providecommand{\dodoi}[1]{doi:~\href{http://doi.org/#1}{\nolinkurl{#1}}}
\providecommand{\doeprint}[1]{\href{http://ascl.net/#1}{\nolinkurl{http://ascl.net/#1}}}
\providecommand{\doarXiv}[1]{\href{https://arxiv.org/abs/#1}{\nolinkurl{https://arxiv.org/abs/#1}}}

\bibitem[{M. Albrow {et~al.}(2000)Albrow, Beaulieu, Caldwell, Dominik, Gaudi,
  Gould, Greenhill, Hill, Kane, Martin, {et~al.}}]{albrow2000detection}
Albrow, M., Beaulieu, J.-P., Caldwell, J., {et~al.} 2000,
  \bibinfo{title}{Detection of rotation in a binary microlens: PLANET
  photometry of MACHO 97-BLG-41,} The Astrophysical Journal, 534, 894

\bibitem[{J.~H. {An}(2005){An}}]{an2005}
{An}, J.~H. 2005, \bibinfo{title}{{Gravitational lens under perturbations:
  symmetry of perturbing potentials with invariant caustics},} \mnras, 356,
  1409, \dodoi{10.1111/j.1365-2966.2004.08581.x}

\bibitem[{J.~H. {An} \& C. {Han}(2002){An} \& {Han}}]{An2002}
{An}, J.~H., \& {Han}, C. 2002, \bibinfo{title}{{Effect of a Wide Binary
  Companion to the Lens on the Astrometric Behavior of Gravitational
  Microlensing Events},} \apj, 573, 351, \dodoi{10.1086/340557}

\bibitem[{D.~P. {Bennett} {et~al.}(2007){Bennett}, {Anderson}, \&
  {Gaudi}}]{bennett2007}
{Bennett}, D.~P., {Anderson}, J., \& {Gaudi}, B.~S. 2007,
  \bibinfo{title}{{Characterization of Gravitational Microlensing Planetary
  Host Stars},} \apj, 660, 781, \dodoi{10.1086/513013}

\bibitem[{V. Bozza(2010)Bozza}]{bozza2010}
Bozza, V. 2010, \bibinfo{title}{Microlensing with an Advanced Contour
  Integration Algorithm: {{Green}}'s Theorem to Third Order, Error Control,
  Optimal Sampling and Limb Darkening: {{Advanced}} Contour Integration in
  Microlensing,} \mnras, 408, 2188, \dodoi{10.1111/j.1365-2966.2010.17265.x}

\bibitem[{V. {Bozza}(2024){Bozza}}]{bozza2024}
{Bozza}, V. 2024, \bibinfo{title}{{RTModel: A platform for real-time modeling
  and massive analyses of microlensing events},} \aap, 688, A83,
  \dodoi{10.1051/0004-6361/202450450}

\bibitem[{V. {Bozza} {et~al.}(2025){Bozza}, {Saggese}, {Covone}, {Rota}, \&
  {Zhang}}]{bozza2025}
{Bozza}, V., {Saggese}, V., {Covone}, G., {Rota}, P., \& {Zhang}, J. 2025,
  \bibinfo{title}{{VBMicroLensing: Three algorithms for multiple lensing with
  contour integration},} \aap, 694, A219, \dodoi{10.1051/0004-6361/202452648}

\bibitem[{J. Bradbury {et~al.}(2018)Bradbury, Frostig, Hawkins, Johnson, Leary,
  Maclaurin, Necula, Paszke, Vander{P}las, Wanderman-{M}ilne, \&
  Zhang}]{jax2018github}
Bradbury, J., Frostig, R., Hawkins, P., {et~al.} 2018, {JAX}: composable
  transformations of {P}ython+{N}um{P}y programs, 0.3.13
  \url{http://github.com/google/jax}

\bibitem[{M.~A. Branch {et~al.}(1999)Branch, Coleman, \&
  Li}]{branch1999subspace}
Branch, M.~A., Coleman, T.~F., \& Li, Y. 1999, \bibinfo{title}{A subspace,
  interior, and conjugate gradient method for large-scale bound-constrained
  minimization problems,} SIAM Journal on Scientific Computing, 21, 1

\bibitem[{S. {Calchi Novati} {et~al.}(2008){Calchi Novati}, {de Luca},
  {Jetzer}, {Mancini}, \& {Scarpetta}}]{CalchiNovati2008}
{Calchi Novati}, S., {de Luca}, F., {Jetzer}, P., {Mancini}, L., \&
  {Scarpetta}, G. 2008, \bibinfo{title}{{Microlensing constraints on the
  Galactic bulge initial mass function},} \aap, 480, 723,
  \dodoi{10.1051/0004-6361:20078439}

\bibitem[{S.-J. {Chung} {et~al.}(2005){Chung}, {Han}, {Park}, {Kim}, {Kang},
  {Ryu}, {Kim}, {Jeon}, {Lee}, {Chang}, {Lee}, \& {Kang}}]{chung2005}
{Chung}, S.-J., {Han}, C., {Park}, B.-G., {et~al.} 2005,
  \bibinfo{title}{{Properties of Central Caustics in Planetary Microlensing},}
  \apj, 630, 535, \dodoi{10.1086/432048}

\bibitem[{K. Cranmer {et~al.}(2020)Cranmer, Brehmer, \&
  Louppe}]{cranmer2020frontier}
Cranmer, K., Brehmer, J., \& Louppe, G. 2020, \bibinfo{title}{The frontier of
  simulation-based inference,} Proceedings of the National Academy of Sciences,
  117, 30055

\bibitem[{ {CSST Collaboration} {et~al.}(2026){CSST Collaboration}, {Gong},
  {Miao}, {Zhan}, {Li}, {Shangguan}, {Li}, {Liu}, {Chen}, {Yuan}, {Zhou},
  {Liu}, {Yu}, {Ji}, {Qi}, {Liu}, {Dai}, {Wang}, {Zheng}, {Hao}, {Dou}, {Ao},
  {Lin}, {Zhang}, {Wang}, {Sun}, {Li}, {Li}, {Xu}, {Li}, {Li}, {Wu}, {Zhang},
  {Wang}, {Bai}, {Cai}, {Cai}, {Cao}, {Chan}, {Chang}, {Chen}, {Chen}, {Chen},
  {Chen}, {Cui}, {Dong}, {Du}, {Duan}, {Fan}, {Fan}, {Fan}, {Fan}, {Fang},
  {Fu}, {Fu}, {Fu}, {Gao}, {Gu}, {Gu}, {Guo}, {Han}, {Hu}, {Huang}, {Ho},
  {Jiang}, {Jiang}, {Jing}, {Kang}, {Kong}, {Li}, {Li}, {Li}, {Li}, {Li}, {Li},
  {Liao}, {Lin}, {Liu}, {Liu}, {Liu}, {Liu}, {Mao}, {Mao}, {Meng}, {Pang},
  {Peng}, {Peng}, {Shan}, {Shen}, {Shen}, {Shen}, {Shi}, {Shi}, {Tan}, {Tian},
  {Wang}, {Wang}, {Wang}, {Wang}, {Wu}, {Wu}, {Wu}, {Xu}, {Xue}, {Xue}, {Yang},
  {Yang}, {Yao}, {Yuan}, {Yuan}, {Zhang}, {Zhang}, {Zhang}, {Zhang}, {Zhang},
  {Zhao}, {Zhao}, {Zhong}, {Zhong}, {Zhou}, {Zhu}, \& {Zu}}]{CSST2026}
{CSST Collaboration}, {Gong}, Y., {Miao}, H., {et~al.} 2026,
  \bibinfo{title}{{Introduction to the Chinese Space Station Survey Telescope
  (CSST)},} Science China Physics, Mechanics, and Astronomy, 69, 239501,
  \dodoi{10.1007/s11433-025-2809-0}

\bibitem[{ DeepMind {et~al.}(2020)DeepMind, Babuschkin, Baumli, Bell,
  Bhupatiraju, Bruce, Buchlovsky, Budden, Cai, Clark, Danihelka, Dedieu,
  Fantacci, Godwin, Jones, Hemsley, Hennigan, Hessel, Hou, Kapturowski, Keck,
  Kemaev, King, Kunesch, Martens, Merzic, Mikulik, Norman, Papamakarios, Quan,
  Ring, Ruiz, Sanchez, Sartran, Schneider, Sezener, Spencer, Srinivasan,
  Stanojevi\'{c}, Stokowiec, Wang, Zhou, \& Viola}]{deepmind2020jax}
DeepMind, Babuschkin, I., Baumli, K., {et~al.} 2020, The {D}eep{M}ind {JAX}
  {E}cosystem, \url{http://github.com/google-deepmind}

\bibitem[{M. Deistler {et~al.}(2022)Deistler, Goncalves, \&
  Macke}]{deistler2022truncated}
Deistler, M., Goncalves, P.~J., \& Macke, J.~H. 2022, \bibinfo{title}{Truncated
  proposals for scalable and hassle-free simulation-based inference,} Advances
  in neural information processing systems, 35, 23135

\bibitem[{M. Deistler {et~al.}(2025)Deistler, Boelts, Steinbach, Moss, Moreau,
  Gloeckler, Rodrigues, Linhart, Lappalainen, Miller,
  {et~al.}}]{deistler2025simulation}
Deistler, M., Boelts, J., Steinbach, P., {et~al.} 2025,
  \bibinfo{title}{Simulation-based inference: A practical guide,} arXiv
  preprint arXiv:2508.12939

\bibitem[{R. {Di Stefano} \& S. {Mao}(1996){Di Stefano} \& {Mao}}]{Stefano1996}
{Di Stefano}, R., \& {Mao}, S. 1996, \bibinfo{title}{{Do Microlensing Events
  Repeat?},} \apj, 457, 93, \dodoi{10.1086/176713}

\bibitem[{M. {Dominik}(1998){Dominik}}]{dominik1997galactic}
{Dominik}, M. 1998, \bibinfo{title}{{Galactic microlensing with rotating
  binaries},} \aap, 329, 361, \dodoi{10.48550/arXiv.astro-ph/9702039}

\bibitem[{M. {Dominik}(1999){Dominik}}]{Dominik1999}
{Dominik}, M. 1999, \bibinfo{title}{{The binary gravitational lens and its
  extreme cases},} \aap, 349, 108, \dodoi{10.48550/arXiv.astro-ph/9903014}

\bibitem[{M. Droettboom {et~al.}(2016)Droettboom, Hunter, Caswell, Firing,
  Nielsen, Elson, Root, Dale, Lee, Sepp{\"a}nen, McDougall, Straw, May,
  Varoquaux, Yu, Ma, Moad, Silvester, Gohlke, W{\"u}rtz, Hisch, Ariza,
  Cimarron, Thomas, Evans, Ivanov, Whitaker, Hobson, mdehoon, \&
  Giuca}]{matplotlib_2016}
Droettboom, M., Hunter, J., Caswell, T.~A., {et~al.} 2016, matplotlib:
  matplotlib v1.5.1, v1.5.1 Zenodo, \dodoi{10.5281/zenodo.44579}

\bibitem[{C. Durkan {et~al.}(2019)Durkan, Bekasov, Murray, \&
  Papamakarios}]{durkan2019neural}
Durkan, C., Bekasov, A., Murray, I., \& Papamakarios, G. 2019,
  \bibinfo{title}{Neural spline flows,} Advances in neural information
  processing systems, 32

\bibitem[{D. Foreman-Mackey(2016)Foreman-Mackey}]{corner}
Foreman-Mackey, D. 2016, \bibinfo{title}{corner.py: Scatterplot matrices in
  Python,} The Journal of Open Source Software, 1, 24,
  \dodoi{10.21105/joss.00024}

\bibitem[{D. {Foreman-Mackey} {et~al.}(2013){Foreman-Mackey}, {Hogg}, {Lang},
  \& {Goodman}}]{foreman-mackey2013}
{Foreman-Mackey}, D., {Hogg}, D.~W., {Lang}, D., \& {Goodman}, J. 2013,
  \bibinfo{title}{{emcee: The MCMC Hammer},} \pasp, 125, 306,
  \dodoi{10.1086/670067}

\bibitem[{F. Gao \& L. Han(2012)Gao \&
  Han}]{gaoImplementingNelderMeadSimplex2012}
Gao, F., \& Han, L. 2012, \bibinfo{title}{Implementing the {{Nelder-Mead}}
  Simplex Algorithm with Adaptive Parameters,} Computational Optimization and
  Applications, 51, 259, \dodoi{10.1007/s10589-010-9329-3}

\bibitem[{B.~S. {Gaudi}(2012){Gaudi}}]{Gaudi2012}
{Gaudi}, B.~S. 2012, \bibinfo{title}{{Microlensing Surveys for Exoplanets},}
  \araa, 50, 411, \dodoi{10.1146/annurev-astro-081811-125518}

\bibitem[{B.~S. {Gaudi} \& A. {Gould}(1997){Gaudi} \&
  {Gould}}]{gaudi&gould1997}
{Gaudi}, B.~S., \& {Gould}, A. 1997, \bibinfo{title}{{Planet Parameters in
  Microlensing Events},} \apj, 486, 85, \dodoi{10.1086/304491}

\bibitem[{J. {Ge} {et~al.}(2022){Ge}, {Zhang}, {Zang}, {Deng}, {Mao}, {Xie},
  {Liu}, {Zhou}, {Willis}, {Huang}, {Howell}, {Feng}, {Zhu}, {Yao}, {Liu},
  {Aizawa}, {Zhu}, {Li}, {Ma}, {Ye}, {Yu}, {Xiang}, {Yu}, {Liu}, {Yang},
  {Wang}, {Shi}, {Fang}, {Zong}, {Liu}, {Zhang}, {Zhang}, {El-Badry}, {Shen},
  {Tam}, {Hu}, {Yang}, {Zou}, {Wu}, {Lei}, {Wei}, {Wu}, {Sun}, {Wang}, {Zhang},
  {Xu}, {Yang}, {Li}, {Xiang}, {Wang}, {Wang}, {Zhang}, {Jia}, {Yuan}, {Zhang},
  {Xuesong Wang}, {Gan}, {Wang}, {Zhao}, {Liu}, {Wei}, {Kang}, {Yang}, {Qi},
  {Liu}, {Zhang}, {Zhu}, {Zhou}, {Zhang}, {Yu}, {Zhang}, {Li}, {Tang}, {Wang},
  {Wang}, {Li}, {Cheng}, {Shen}, {Li}, {Pan}, {Yang}, {Gao}, {Song}, {Wang},
  {Zhang}, {Chen}, {Wang}, {Zhang}, {Wang}, {Zeng}, {Zheng}, {Zhu}, {Guo},
  {Zhang}, {Li}, {Wen}, {Feng}, {Chen}, {Chen}, {Han}, {Yang}, {Wang}, {Duan},
  {Huang}, {Liang}, {Bi}, {Gai}, {Ge}, {Guo}, {Huang}, {Li}, {Li}, {Li},
  {Yuxi}, {Lu}, {Rix}, {Shi}, {Song}, {Tang}, {Ting}, {Wu}, {Wu}, {Yang},
  {Yin}, {Gould}, {Lee}, {Dong}, {Yee}, {Shvartzvald}, {Yang}, {Kuang},
  {Zhang}, {Liao}, {Qi}, {Yang}, {Zhang}, {Jiang}, {Ou}, {Li}, {Beck},
  {Bedding}, {Campante}, {Chaplin}, {Christensen-Dalsgaard}, {Garc{\'\i}a},
  {Gaulme}, {Gizon}, {Hekker}, {Huber}, {Khanna}, {Li}, {Mathur}, {Miglio},
  {Mosser}, {Ong}, {Santos}, {Stello}, {Bowman}, {Lares-Martiz}, {Murphy},
  {Niu}, {Ma}, {Moln{\'a}r}, {Fu}, {De Cat}, {Su}, \& {consortium}}]{Ge2022}
{Ge}, J., {Zhang}, H., {Zang}, W., {et~al.} 2022, \bibinfo{title}{{ET White
  Paper: To Find the First Earth 2.0},} arXiv e-prints, arXiv:2206.06693,
  \dodoi{10.48550/arXiv.2206.06693}

\bibitem[{A. {Gould}(1992){Gould}}]{Gould1992}
{Gould}, A. 1992, \bibinfo{title}{{Extending the MACHO Search to approximately
  10 6 M sub sun},} \apj, 392, 442, \dodoi{10.1086/171443}

\bibitem[{A. {Gould}(2000){Gould}}]{Gould2000}
{Gould}, A. 2000, \bibinfo{title}{{A Natural Formalism for Microlensing},}
  \apj, 542, 785, \dodoi{10.1086/317037}

\bibitem[{A. {Gould} \& A. {Loeb}(1992){Gould} \& {Loeb}}]{Gould&Loeb1992}
{Gould}, A., \& {Loeb}, A. 1992, \bibinfo{title}{{Discovering Planetary Systems
  through Gravitational Microlenses},} \apj, 396, 104, \dodoi{10.1086/171700}

\bibitem[{A. {Gould} {et~al.}(2021){Gould}, {Zang}, {Mao}, \& {Dong}}]{CMST}
{Gould}, A., {Zang}, W.-C., {Mao}, S., \& {Dong}, S.-B. 2021,
  \bibinfo{title}{{Masses for free-floating planets and dwarf planets},}
  Research in Astronomy and Astrophysics, 21, 133,
  \dodoi{10.1088/1674-4527/21/6/133}

\bibitem[{D. Greenberg {et~al.}(2019)Greenberg, Nonnenmacher, \&
  Macke}]{greenberg2019automatic}
Greenberg, D., Nonnenmacher, M., \& Macke, J. 2019, \bibinfo{title}{Automatic
  posterior transformation for likelihood-free inference,} in International
  conference on machine learning, PMLR, 2404--2414

\bibitem[{K. {Griest} \& N. {Safizadeh}(1998){Griest} \&
  {Safizadeh}}]{Griest1998}
{Griest}, K., \& {Safizadeh}, N. 1998, \bibinfo{title}{{The Use of
  High-Magnification Microlensing Events in Discovering Extrasolar Planets},}
  \apj, 500, 37, \dodoi{10.1086/305729}

\bibitem[{F.~R. Hampel(1974)Hampel}]{hampel1974influence}
Hampel, F.~R. 1974, \bibinfo{title}{The influence curve and its role in robust
  estimation,} Journal of the american statistical association, 69, 383

\bibitem[{C. {Han}(2006){Han}}]{Han2006}
{Han}, C. 2006, \bibinfo{title}{{Properties of Planetary Caustics in
  Gravitational Microlensing},} \apj, 638, 1080, \dodoi{10.1086/498937}

\bibitem[{C. {Han} {et~al.}(2026){Han}, {Udalski}, {Bond}, {Lee}, {Albrow},
  {Chung}, {Gould}, {Jung}, {Hwang}, {Ryu}, {Shvartzvald}, {Shin}, {Yee},
  {Zang}, {Yang}, {Kim}, {Kim}, {Kim}, {Lee}, {Cha}, {Lee}, {Park}, {Pogge},
  {Mr{\'o}z}, {Szyma{\'n}ski}, {Skowron}, {Poleski}, {Soszy{\'n}ski},
  {Pietrukowicz}, {Koz{\l}owski}, {Rybicki}, {Iwanek}, {Ulaczyk}, {Wrona},
  {Gromadzki}, {Mr{\'o}z}, {Abe}, {Bennett}, {Bhattacharya}, {Hamada}, {Hirao},
  {Idei}, {Ishitani Silva}, {Makida}, {Miyazaki}, {Muraki}, {Nagai}, {Nagano},
  {Nakayama}, {Nishio}, {Nunota}, {Ogawa}, {Oishi}, {Okumoto}, {Olmschenk},
  {Ranc}, {Rattenbury}, {Satoh}, {Sumi}, {Suzuki}, {Tamaoki}, {Terry},
  {Tristram}, {Vandorou}, \& {Yama}}]{Han2026}
{Han}, C., {Udalski}, A., {Bond}, I.~A., {et~al.} 2026,
  \bibinfo{title}{{Candidate Microlensing Brown Dwarfs in Binary Lens Systems
  from the 2023--2025 Observing Seasons},} arXiv e-prints, arXiv:2604.07932,
  \dodoi{10.48550/arXiv.2604.07932}

\bibitem[{C.~R. {Harris} {et~al.}(2020){Harris}, {Millman}, {van der Walt},
  {Gommers}, {Virtanen}, {Cournapeau}, {Wieser}, {Taylor}, {Berg}, {Smith},
  {Kern}, {Picus}, {Hoyer}, {van Kerkwijk}, {Brett}, {Haldane}, {del R{\'\i}o},
  {Wiebe}, {Peterson}, {G{\'e}rard-Marchant}, {Sheppard}, {Reddy}, {Weckesser},
  {Abbasi}, {Gohlke}, \& {Oliphant}}]{numpy2020}
{Harris}, C.~R., {Millman}, K.~J., {van der Walt}, S.~J., {et~al.} 2020,
  \bibinfo{title}{{Array programming with NumPy},} \nat, 585, 357,
  \dodoi{10.1038/s41586-020-2649-2}

\bibitem[{K. He {et~al.}(2016{\natexlab{a}})He, Zhang, Ren, \&
  Sun}]{he2016deep}
He, K., Zhang, X., Ren, S., \& Sun, J. 2016{\natexlab{a}}, \bibinfo{title}{Deep
  residual learning for image recognition,} in Proceedings of the IEEE
  conference on computer vision and pattern recognition, 770--778

\bibitem[{K. He {et~al.}(2016{\natexlab{b}})He, Zhang, Ren, \&
  Sun}]{he2016identity}
He, K., Zhang, X., Ren, S., \& Sun, J. 2016{\natexlab{b}},
  \bibinfo{title}{Identity mappings in deep residual networks,} in European
  conference on computer vision, Springer, 630--645

\bibitem[{J.~D. Hunter(2007)Hunter}]{matplotlib2007}
Hunter, J.~D. 2007, \bibinfo{title}{Matplotlib: A 2D Graphics Environment,}
  Computing in Science \& Engineering, 9, 90, \dodoi{10.1109/MCSE.2007.55}

\bibitem[{Y.~K. {Jung} {et~al.}(2022){Jung}, {Zang}, {Han}, {Gould}, {Udalski},
  {Albrow}, {Chung}, {Hwang}, {Ryu}, {Shin}, {Shvartzvald}, {Yang}, {Yee},
  {Cha}, {Kim}, {Kim}, {Lee}, {Lee}, {Lee}, {Park}, {Pogge}, {KMTNet
  Collaboration}, {Mr{\'o}z}, {Szyma{\'n}ski}, {Skowron}, {Poleski},
  {Soszy{\'n}ski}, {Pietrukowicz}, {Koz{\l}owski}, {Ulaczyk}, {Rybicki},
  {Iwanek}, {Wrona}, \& {OGLE Collaboration}}]{Jung2022}
{Jung}, Y.~K., {Zang}, W., {Han}, C., {et~al.} 2022,
  \bibinfo{title}{{Systematic KMTNet Planetary Anomaly Search. VI. Complete
  Sample of 2018 Sub-prime-field Planets},} \aj, 164, 262,
  \dodoi{10.3847/1538-3881/ac9c5c}

\bibitem[{P. {Kidger}(2022){Kidger}}]{Kidger2022}
{Kidger}, P. 2022, \bibinfo{title}{{On Neural Differential Equations},} arXiv
  e-prints, arXiv:2202.02435, \dodoi{10.48550/arXiv.2202.02435}

\bibitem[{P. Kidger \& C. Garcia(2021)Kidger \& Garcia}]{kidger2021equinox}
Kidger, P., \& Garcia, C. 2021, \bibinfo{title}{{E}quinox: neural networks in
  {JAX} via callable {P}y{T}rees and filtered transformations,} Differentiable
  Programming workshop at Neural Information Processing Systems 2021

\bibitem[{P. Kidger \& T. Lyons(2021)Kidger \& Lyons}]{kidger2021signatory}
Kidger, P., \& Lyons, T. 2021, \bibinfo{title}{{S}ignatory: differentiable
  computations of the signature and logsignature transforms, on both {CPU} and
  {GPU},} in International Conference on Learning Representations

\bibitem[{P. Kidger {et~al.}(2020)Kidger, Morrill, Foster, \&
  Lyons}]{kidger2020}
Kidger, P., Morrill, J., Foster, J., \& Lyons, T. 2020, \bibinfo{title}{Neural
  controlled differential equations for irregular time series,} Advances in
  neural information processing systems, 33, 6696

\bibitem[{S.-L. {Kim} {et~al.}(2016){Kim}, {Lee}, {Park}, {Kim}, {Cha}, {Lee},
  {Han}, {Chun}, \& {Yuk}}]{kmtnet2016}
{Kim}, S.-L., {Lee}, C.-U., {Park}, B.-G., {et~al.} 2016,
  \bibinfo{title}{{KMTNET: A Network of 1.6 m Wide-Field Optical Telescopes
  Installed at Three Southern Observatories},} Journal of Korean Astronomical
  Society, 49, 37, \dodoi{10.5303/JKAS.2016.49.1.37}

\bibitem[{D.~P. Kingma {et~al.}(2016)Kingma, Salimans, Jozefowicz, Chen,
  Sutskever, \& Welling}]{kingma2016improved}
Kingma, D.~P., Salimans, T., Jozefowicz, R., {et~al.} 2016,
  \bibinfo{title}{Improved variational inference with inverse autoregressive
  flow,} Advances in neural information processing systems, 29

\bibitem[{T. Kluyver {et~al.}(2016)Kluyver, Ragan-Kelley, P{\'e}rez, Granger,
  Bussonnier, Frederic, Kelley, Hamrick, Grout, Corlay, Ivanov, Whitaker,
  Hobson, Avila, Abdalla, Willing, \& development team}]{jupyter}
Kluyver, T., Ragan-Kelley, B., P{\'e}rez, F., {et~al.} 2016,
  \bibinfo{title}{Jupyter Notebooks -- a publishing format for reproducible
  computational workflows,} in Positioning and Power in Academic Publishing:
  Players, Agents and Agendas, ed. F.~Loizides \& B.~Scmidt (IOS Press),
  87--90.
\newblock \url{https://eprints.soton.ac.uk/403913/}

\bibitem[{C. {Liebig} {et~al.}(2015){Liebig}, {D'Ago}, {Bozza}, \&
  {Dominik}}]{Liebig2015}
{Liebig}, C., {D'Ago}, G., {Bozza}, V., \& {Dominik}, M. 2015,
  \bibinfo{title}{{The complete catalogue of light curves in equal-mass binary
  microlensing},} \mnras, 450, 1565, \dodoi{10.1093/mnras/stv733}

\bibitem[{I. Loshchilov \& F. Hutter(2019)Loshchilov \&
  Hutter}]{Loshchilov2019}
Loshchilov, I., \& Hutter, F. 2019, \bibinfo{title}{Decoupled Weight Decay
  Regularization,} in 7th International Conference on Learning Representations,
  {ICLR} 2019, New Orleans, LA, USA, May 6-9, 2019 (OpenReview.net).
\newblock \url{https://openreview.net/forum?id=Bkg6RiCqY7}

\bibitem[{J.-M. Lueckmann {et~al.}(2017)Lueckmann, Goncalves, Bassetto,
  {\"O}cal, Nonnenmacher, \& Macke}]{lueckmann2017flexible}
Lueckmann, J.-M., Goncalves, P.~J., Bassetto, G., {et~al.} 2017,
  \bibinfo{title}{Flexible statistical inference for mechanistic models of
  neural dynamics,} Advances in neural information processing systems, 30

\bibitem[{R.~H. {Lupton} {et~al.}(1999){Lupton}, {Gunn}, \&
  {Szalay}}]{lupton1999}
{Lupton}, R.~H., {Gunn}, J.~E., \& {Szalay}, A.~S. 1999, \bibinfo{title}{{A
  Modified Magnitude System that Produces Well-Behaved Magnitudes, Colors, and
  Errors Even for Low Signal-to-Noise Ratio Measurements},} \aj, 118, 1406,
  \dodoi{10.1086/301004}

\bibitem[{S. Mao \& B. Paczynski(1991)Mao \& Paczynski}]{mao1991}
Mao, S., \& Paczynski, B. 1991, \bibinfo{title}{Gravitational {{Microlensing}}
  by {{Double Stars}} and {{Planetary Systems}},} \apj, 374, L37,
  \dodoi{10.1086/186066}

\bibitem[{J. {Morrill} {et~al.}(2020){Morrill}, {Fermanian}, {Kidger}, \&
  {Lyons}}]{Morrill2020signature_method}
{Morrill}, J., {Fermanian}, A., {Kidger}, P., \& {Lyons}, T. 2020,
  \bibinfo{title}{{A Generalised Signature Method for Multivariate Time Series
  Feature Extraction},} arXiv e-prints, arXiv:2006.00873,
  \dodoi{10.48550/arXiv.2006.00873}

\bibitem[{J. {Morrill} {et~al.}(2021){Morrill}, {Kidger}, {Yang}, \&
  {Lyons}}]{Morrill2021Hermite_cubic_splines}
{Morrill}, J., {Kidger}, P., {Yang}, L., \& {Lyons}, T. 2021,
  \bibinfo{title}{{Neural Controlled Differential Equations for Online
  Prediction Tasks},} arXiv e-prints, arXiv:2106.11028,
  \dodoi{10.48550/arXiv.2106.11028}

\bibitem[{J. Morrill {et~al.}(2021)Morrill, Salvi, Kidger, \&
  Foster}]{Morrill2021neuralRDE}
Morrill, J., Salvi, C., Kidger, P., \& Foster, J. 2021, \bibinfo{title}{Neural
  Rough Differential Equations for Long Time Series,} in Proceedings of Machine
  Learning Research, Vol. 139, Proceedings of the 38th International Conference
  on Machine Learning, ed. M.~Meila \& T.~Zhang (PMLR), 7829--7838.
\newblock \url{https://proceedings.mlr.press/v139/morrill21b.html}

\bibitem[{P. {Mr{\'o}z} {et~al.}(2017){Mr{\'o}z}, {Udalski}, {Skowron},
  {Poleski}, {Koz{\l}owski}, {Szyma{\'n}ski}, {Soszy{\'n}ski}, {Wyrzykowski},
  {Pietrukowicz}, {Ulaczyk}, {Skowron}, \& {Pawlak}}]{mroz2017}
{Mr{\'o}z}, P., {Udalski}, A., {Skowron}, J., {et~al.} 2017,
  \bibinfo{title}{{No large population of unbound or wide-orbit Jupiter-mass
  planets},} \nat, 548, 183, \dodoi{10.1038/nature23276}

\bibitem[{P. {Mr{\'o}z} {et~al.}(2019){Mr{\'o}z}, {Udalski}, {Skowron},
  {Szyma{\'n}ski}, {Soszy{\'n}ski}, {Wyrzykowski}, {Pietrukowicz},
  {Koz{\l}owski}, {Poleski}, {Ulaczyk}, {Rybicki}, \& {Iwanek}}]{mroz2019}
{Mr{\'o}z}, P., {Udalski}, A., {Skowron}, J., {et~al.} 2019,
  \bibinfo{title}{{Microlensing Optical Depth and Event Rate toward the
  Galactic Bulge from 8 yr of OGLE-IV Observations},} \apjs, 244, 29,
  \dodoi{10.3847/1538-4365/ab426b}

\bibitem[{J.~A. Nelder \& R. Mead(1965)Nelder \&
  Mead}]{nelderSimplexMethodFunction1965}
Nelder, J.~A., \& Mead, R. 1965, \bibinfo{title}{A {{Simplex Method}} for
  {{Function Minimization}},} The Computer Journal, 7, 308,
  \dodoi{10.1093/comjnl/7.4.308}

\bibitem[{R.~A.~P. {Oliveira} {et~al.}(2025){Oliveira}, {Poleski}, {Mr{\'o}z},
  {Udalski}, {Skowron}, {Mr{\'o}z}, {Szyma{\'n}ski}, {Soszy{\'n}ski},
  {Ulaczyk}, {Pietrukowicz}, {Rybicki}, {Iwanek}, {Wrona}, \&
  {Gromadzki}}]{Oliveira2025}
{Oliveira}, R.~A.~P., {Poleski}, R., {Mr{\'o}z}, P., {et~al.} 2025,
  \bibinfo{title}{{Automated Detection and Modeling of Binary Microlensing
  Events in OGLE-IV data. I. Events with Well-Separated Bumps},} \actaa, 75,
  27, \dodoi{10.32023/0001-5237/75.1.2}

\bibitem[{B. {Paczynski}(1986){Paczynski}}]{paczynski1986}
{Paczynski}, B. 1986, \bibinfo{title}{{Gravitational Microlensing by the
  Galactic Halo},} \apj, 304, 1, \dodoi{10.1086/164140}

\bibitem[{G. Papamakarios \& I. Murray(2016)Papamakarios \&
  Murray}]{papamakarios2016fast}
Papamakarios, G., \& Murray, I. 2016, \bibinfo{title}{Fast $\varepsilon$-free
  inference of simulation models with bayesian conditional density estimation,}
  Advances in neural information processing systems, 29

\bibitem[{G. Papamakarios {et~al.}(2021)Papamakarios, Nalisnick, Rezende,
  Mohamed, \& Lakshminarayanan}]{papamakarios2021normalizing}
Papamakarios, G., Nalisnick, E., Rezende, D.~J., Mohamed, S., \&
  Lakshminarayanan, B. 2021, \bibinfo{title}{Normalizing flows for
  probabilistic modeling and inference,} Journal of Machine Learning Research,
  22, 1

\bibitem[{G. Papamakarios {et~al.}(2017)Papamakarios, Pavlakou, \&
  Murray}]{papamakarios2017masked}
Papamakarios, G., Pavlakou, T., \& Murray, I. 2017, \bibinfo{title}{Masked
  autoregressive flow for density estimation,} Advances in neural information
  processing systems, 30

\bibitem[{R. Pascanu {et~al.}(2013)Pascanu, Mikolov, \&
  Bengio}]{pascanu2013difficulty}
Pascanu, R., Mikolov, T., \& Bengio, Y. 2013, \bibinfo{title}{On the difficulty
  of training recurrent neural networks,} in International conference on
  machine learning, Pmlr, 1310--1318

\bibitem[{M.~T. {Penny} {et~al.}(2019){Penny}, {Gaudi}, {Kerins}, {Rattenbury},
  {Mao}, {Robin}, \& {Calchi Novati}}]{penny2019}
{Penny}, M.~T., {Gaudi}, B.~S., {Kerins}, E., {et~al.} 2019,
  \bibinfo{title}{{Predictions of the WFIRST Microlensing Survey. I. Bound
  Planet Detection Rates},} \apjs, 241, 3, \dodoi{10.3847/1538-4365/aafb69}

\bibitem[{H. {Ren} \& W. {Zhu}(2025){Ren} \& {Zhu}}]{ren2025}
{Ren}, H., \& {Zhu}, W. 2025, \bibinfo{title}{{A Differentiable Binary
  Microlensing Model Using Adaptive Contour Integration Method},} \aj, 169,
  170, \dodoi{10.3847/1538-3881/adb1b2}

\bibitem[{D. Rezende \& S. Mohamed(2015)Rezende \&
  Mohamed}]{rezende2015variational}
Rezende, D., \& Mohamed, S. 2015, \bibinfo{title}{Variational inference with
  normalizing flows,} in International conference on machine learning, PMLR,
  1530--1538

\bibitem[{T. {Sako} {et~al.}(2008){Sako}, {Sekiguchi}, {Sasaki}, {Okajima},
  {Abe}, {Bond}, {Hearnshaw}, {Itow}, {Kamiya}, {Kilmartin}, {Masuda},
  {Matsubara}, {Muraki}, {Rattenbury}, {Sullivan}, {Sumi}, {Tristram},
  {Yanagisawa}, \& {Yock}}]{Sako2008}
{Sako}, T., {Sekiguchi}, T., {Sasaki}, M., {et~al.} 2008,
  \bibinfo{title}{{MOA-cam3: a wide-field mosaic CCD camera for a gravitational
  microlensing survey in New Zealand},} Experimental Astronomy, 22, 51,
  \dodoi{10.1007/s10686-007-9082-5}

\bibitem[{A. Savitzky \& M.~J. Golay(1964)Savitzky \&
  Golay}]{savitzky1964smoothing}
Savitzky, A., \& Golay, M.~J. 1964, \bibinfo{title}{Smoothing and
  differentiation of data by simplified least squares procedures.,} Analytical
  chemistry, 36, 1627

\bibitem[{Y. {Shvartzvald} {et~al.}(2016){Shvartzvald}, {Maoz}, {Udalski},
  {Sumi}, {Friedmann}, {Kaspi}, {Poleski}, {Szyma{\'n}ski}, {Skowron},
  {Koz{\l}owski}, {Wyrzykowski}, {Mr{\'o}z}, {Pietrukowicz}, {Pietrzy{\'n}ski},
  {Soszy{\'n}ski}, {Ulaczyk}, {Abe}, {Barry}, {Bennett}, {Bhattacharya},
  {Bond}, {Freeman}, {Inayama}, {Itow}, {Koshimoto}, {Ling}, {Masuda}, {Fukui},
  {Matsubara}, {Muraki}, {Ohnishi}, {Rattenbury}, {Saito}, {Sullivan},
  {Suzuki}, {Tristram}, {Wakiyama}, \& {Yonehara}}]{Wise}
{Shvartzvald}, Y., {Maoz}, D., {Udalski}, A., {et~al.} 2016,
  \bibinfo{title}{{The frequency of snowline-region planets from four years of
  OGLE-MOA-Wise second-generation microlensing},} \mnras, 457, 4089,
  \dodoi{10.1093/mnras/stw191}

\bibitem[{J. Skowron {et~al.}(2011)Skowron, Udalski, Gould, Dong, Monard, Han,
  Nelson, McCormick, Moorhouse, Thornley, {et~al.}}]{skowron2011binary}
Skowron, J., Udalski, A., Gould, A., {et~al.} 2011, \bibinfo{title}{Binary
  microlensing event OGLE-2009-BLG-020 gives verifiable mass, distance, and
  orbit predictions,} The Astrophysical Journal, 738, 87

\bibitem[{N. {Smyth} {et~al.}(2025){Smyth}, {Perreault-Levasseur}, \&
  {Hezaveh}}]{Smyth2025_Transformer_Embeddings}
{Smyth}, N., {Perreault-Levasseur}, L., \& {Hezaveh}, Y. 2025,
  \bibinfo{title}{{Transformer Embeddings for Fast Microlensing Inference},}
  arXiv e-prints, arXiv:2512.11687, \dodoi{10.48550/arXiv.2512.11687}

\bibitem[{D. Spergel {et~al.}(2015)Spergel, Gehrels, Baltay, Bennett,
  Breckinridge, Donahue, Dressler, Gaudi, Greene, Guyon,
  {et~al.}}]{spergel2015wide}
Spergel, D., Gehrels, N., Baltay, C., {et~al.} 2015, \bibinfo{title}{Wide-field
  infrarred survey telescope-astrophysics focused telescope assets WFIRST-AFTA
  2015 report,} ArXiv e-prints, arXiv

\bibitem[{D. {Suzuki} {et~al.}(2016){Suzuki}, {Bennett}, {Sumi}, {Bond},
  {Rogers}, {Abe}, {Asakura}, {Bhattacharya}, {Donachie}, {Freeman}, {Fukui},
  {Hirao}, {Itow}, {Koshimoto}, {Li}, {Ling}, {Masuda}, {Matsubara}, {Muraki},
  {Nagakane}, {Onishi}, {Oyokawa}, {Rattenbury}, {Saito}, {Sharan}, {Shibai},
  {Sullivan}, {Tristram}, {Yonehara}, \& {MOA Collaboration}}]{Suzuki2016}
{Suzuki}, D., {Bennett}, D.~P., {Sumi}, T., {et~al.} 2016, \bibinfo{title}{{The
  Exoplanet Mass-ratio Function from the MOA-II Survey: Discovery of a Break
  and Likely Peak at a Neptune Mass},} \apj, 833, 145,
  \dodoi{10.3847/1538-4357/833/2/145}

\bibitem[{S. Talts {et~al.}(2018)Talts, Betancourt, Simpson, Vehtari, \&
  Gelman}]{talts2018validating}
Talts, S., Betancourt, M., Simpson, D., Vehtari, A., \& Gelman, A. 2018,
  \bibinfo{title}{Validating Bayesian inference algorithms with
  simulation-based calibration,} arXiv preprint arXiv:1804.06788

\bibitem[{S.~K. {Terry} {et~al.}(2026){Terry}, {Anderson}, {Beichman},
  {Bennett}, {Bhattacharya}, {Beaulieu}, {Gaudi}, {Green}, {Huston}, {Lu},
  {Lucas}, {Nataf}, {Penny}, {Rektsini}, {Rodriguez Sanchez-Vahamonde}, \&
  {Vandorou}}]{Terry2026_HST_Bulge_Survey}
{Terry}, S.~K., {Anderson}, J., {Beichman}, C.~A., {et~al.} 2026,
  \bibinfo{title}{{An HST Wide Field Survey of the Galactic Bulge: Overview,
  Strategy, and First Results},} arXiv e-prints, arXiv:2605.06778.
\newblock \doarXiv{2605.06778}

\bibitem[{A. {Udalski} {et~al.}(2015){Udalski}, {Szyma{\'n}ski}, \&
  {Szyma{\'n}ski}}]{OGLEIV}
{Udalski}, A., {Szyma{\'n}ski}, M.~K., \& {Szyma{\'n}ski}, G. 2015,
  \bibinfo{title}{{OGLE-IV: Fourth Phase of the Optical Gravitational Lensing
  Experiment},} \actaa, 65, 1.
\newblock \doarXiv{1504.05966}

\bibitem[{S. van~der Walt {et~al.}(2011)van~der Walt, Colbert, \&
  Varoquaux}]{numpy2011}
van~der Walt, S., Colbert, S.~C., \& Varoquaux, G. 2011, \bibinfo{title}{The
  NumPy Array: A Structure for Efficient Numerical Computation,} Computing in
  Science \& Engineering, 13, 22, \dodoi{10.1109/MCSE.2011.37}

\bibitem[{P. {Virtanen} {et~al.}(2020){Virtanen}, {Gommers}, {Oliphant},
  {Haberland}, {Reddy}, {Cournapeau}, {Burovski}, {Peterson}, {Weckesser},
  {Bright}, {van der Walt}, {Brett}, {Wilson}, {Millman}, {Mayorov}, {Nelson},
  {Jones}, {Kern}, {Larson}, {Carey}, {Polat}, {Feng}, {Moore}, {VanderPlas},
  {Laxalde}, {Perktold}, {Cimrman}, {Henriksen}, {Quintero}, {Harris},
  {Archibald}, {Ribeiro}, {Pedregosa}, {van Mulbregt}, \& {SciPy 1.0
  Contributors}}]{scipy2020}
{Virtanen}, P., {Gommers}, R., {Oliphant}, T.~E., {et~al.} 2020,
  \bibinfo{title}{{SciPy 1.0: fundamental algorithms for scientific computing
  in Python},} Nature Methods, 17, 261, \dodoi{10.1038/s41592-019-0686-2}

\bibitem[{D. Ward(2025)Ward}]{ward2023flowjax}
Ward, D. 2025, FlowJAX: Distributions and Normalizing Flows in Jax, 17.2.1
  \dodoi{10.5281/zenodo.10402073}

\bibitem[{C. {Wegg} {et~al.}(2017){Wegg}, {Gerhard}, \& {Portail}}]{Wegg2017}
{Wegg}, C., {Gerhard}, O., \& {Portail}, M. 2017, \bibinfo{title}{{The Initial
  Mass Function of the Inner Galaxy Measured from OGLE-III Microlensing
  Timescales},} \apjl, 843, L5, \dodoi{10.3847/2041-8213/aa794e}

\bibitem[{H.~J. {Witt} \& S. {Mao}(1994){Witt} \& {Mao}}]{witt&mao1994}
{Witt}, H.~J., \& {Mao}, S. 1994, \bibinfo{title}{{Can Lensed Stars Be Regarded
  as Pointlike for Microlensing by MACHOs?},} \apj, 430, 505,
  \dodoi{10.1086/174426}

\bibitem[{S. {Yan} \& W. {Zhu}(2022){Yan} \& {Zhu}}]{Yan2022}
{Yan}, S., \& {Zhu}, W. 2022, \bibinfo{title}{{Measuring Microlensing Parallax
  via Simultaneous Observations from Chinese Space Station Telescope and Roman
  Telescope},} Research in Astronomy and Astrophysics, 22, 025006,
  \dodoi{10.1088/1674-4527/ac3c44}

\bibitem[{H. {Yang} {et~al.}(2024){Yang}, {Yee}, {Hwang}, {Qian}, {Bond},
  {Gould}, {Hu}, {Zhang}, {Mao}, {Zhu}, {Albrow}, {Chung}, {Kim}, {Park},
  {Han}, {Jung}, {Ryu}, {Shin}, {Shvartzvald}, {Cha}, {Kim}, {Kim}, {Lee},
  {Lee}, {Lee}, {Pogge}, {Zang}, {Abe}, {Barry}, {Bennett}, {Bhattacharya},
  {Donachie}, {Fujii}, {Fukui}, {Hirao}, {Itow}, {Kirikawa}, {Kondo},
  {Koshimoto}, {Silva}, {Li}, {Matsubara}, {Muraki}, {Suzuki}, {Tristram},
  {Yonehara}, {Ranc}, {Miyazaki}, {Olmschenk}, {Rattenbury}, {Satoh}, {Shoji},
  {Sumi}, {Tanaka}, \& {Yamawaki}}]{yang2024}
{Yang}, H., {Yee}, J.~C., {Hwang}, K.-H., {et~al.} 2024,
  \bibinfo{title}{{Systematic reanalysis of KMTNet microlensing events, paper
  I: Updates of the photometry pipeline and a new planet candidate},} \mnras,
  528, 11, \dodoi{10.1093/mnras/stad3672}

\bibitem[{W. {Zang} {et~al.}(2021){Zang}, {Hwang}, {Udalski}, {Wang}, {Zhu},
  {Sumi}, {Yee}, {Gould}, {Mao}, {Zhang}, {Albrow}, {Chung}, {Han}, {Jung},
  {Ryu}, {Shin}, {Shvartzvald}, {Cha}, {Kim}, {Kim}, {Kim}, {Lee}, {Lee},
  {Lee}, {Park}, {Pogge}, {Mr{\'o}z}, {Skowron}, {Poleski}, {Szyma{\'n}ski},
  {Soszy{\'n}ski}, {Pietrukowicz}, {Koz{\l}owski}, {Ulaczyk}, {Rybicki},
  {Iwanek}, {Wrona}, {Gromadzki}, {Bond}, {Abe}, {Barry}, {Bennett},
  {Bhattacharya}, {Donachie}, {Fujii}, {Fukui}, {Hirao}, {Itow}, {Kirikawa},
  {Kondo}, {Koshimoto}, {Li}, {Matsubara}, {Muraki}, {Miyazaki}, {Olmschenk},
  {Ranc}, {Rattenbury}, {Satoh}, {Shoji}, {Ishitani Silva}, {Suzuki}, {Tanaka},
  {Tristram}, {Yamawaki}, {Yonehara}, {Beichman}, {Bryden}, {Calchi Novati},
  {Carey}, {Gaudi}, {Henderson}, {Johnson}, \& {Spitzer Team}}]{Zang2021}
{Zang}, W., {Hwang}, K.-H., {Udalski}, A., {et~al.} 2021,
  \bibinfo{title}{{Systematic KMTNet Planetary Anomaly Search. I.
  OGLE-2019-BLG-1053Lb, a Buried Terrestrial Planet},} \aj, 162, 163,
  \dodoi{10.3847/1538-3881/ac12d4}

\bibitem[{W. {Zang} {et~al.}(2025){Zang}, {Jung}, {Yee}, {Hwang}, {Yang},
  {Udalski}, {Sumi}, {Gould}, {Mao}, {Albrow}, {Chung}, {Han}, {Ryu}, {Shin},
  {Shvartzvald}, {Cha}, {Kim}, {Kim}, {Kim}, {Lee}, {Lee}, {Lee}, {Park},
  {Pogge}, {Zhang}, {Kuang}, {Wang}, {Zhang}, {Hu}, {Zhu}, {Mr{\'o}z},
  {Skowron}, {Poleski}, {Szyma{\'n}ski}, {Soszy{\'n}ski}, {Pietrukowicz},
  {Koz{\l}owski}, {Ulaczyk}, {Rybicki}, {Iwanek}, {Wrona}, {Gromadzki}, {Abe},
  {Barry}, {Bennett}, {Bhattacharya}, {Bond}, {Fujii}, {Fukui}, {Hamada},
  {Hirao}, {Silva}, {Itow}, {Kirikawa}, {Koshimoto}, {Matsubara}, {Miyazaki},
  {Muraki}, {Olmschenk}, {Ranc}, {Rattenbury}, {Satoh}, {Suzuki}, {Tomoyoshi},
  {Tristram}, {Vandorou}, {Yama}, \& {Yamashita}}]{Zang2025}
{Zang}, W., {Jung}, Y.~K., {Yee}, J.~C., {et~al.} 2025,
  \bibinfo{title}{{Microlensing events indicate that super-Earth exoplanets are
  common in Jupiter-like orbits},} Science, 388, 400,
  \dodoi{10.1126/science.adn6088}

\bibitem[{K. {Zhang} {et~al.}(2021){Zhang}, {Bloom}, {Gaudi}, {Lanusse}, {Lam},
  \& {Lu}}]{zhang2021romanML}
{Zhang}, K., {Bloom}, J.~S., {Gaudi}, B.~S., {et~al.} 2021,
  \bibinfo{title}{{Real-time Likelihood-free Inference of Roman Binary
  Microlensing Events with Amortized Neural Posterior Estimation},} \aj, 161,
  262, \dodoi{10.3847/1538-3881/abf42e}

\bibitem[{K. {Zhang} \& B.~S. {Gaudi}(2022){Zhang} \&
  {Gaudi}}]{Zhang&Gaudi2022}
{Zhang}, K., \& {Gaudi}, B.~S. 2022, \bibinfo{title}{{A Mathematical Treatment
  of the Offset Microlensing Degeneracy},} \apjl, 936, L22,
  \dodoi{10.3847/2041-8213/ac8c2b}

\bibitem[{K. Zhang {et~al.}(2022)Zhang, Gaudi, \& Bloom}]{zhang2022ubiquitous}
Zhang, K., Gaudi, B.~S., \& Bloom, J.~S. 2022, \bibinfo{title}{A ubiquitous
  unifying degeneracy in two-body microlensing systems,} Nature Astronomy, 6,
  782

\bibitem[{H. {Zhao} \& W. {Zhu}(2022){Zhao} \& {Zhu}}]{zhao&zhu2022}
{Zhao}, H., \& {Zhu}, W. 2022, \bibinfo{title}{{MAGIC: Microlensing Analysis
  Guided by Intelligent Computation},} \aj, 164, 192,
  \dodoi{10.3847/1538-3881/ac9230}

\bibitem[{W. {Zhu} {et~al.}(2014){Zhu}, {Penny}, {Mao}, {Gould}, \&
  {Gendron}}]{zhu2014}
{Zhu}, W., {Penny}, M., {Mao}, S., {Gould}, A., \& {Gendron}, R. 2014,
  \bibinfo{title}{{Predictions for Microlensing Planetary Events from Core
  Accretion Theory},} \apj, 788, 73, \dodoi{10.1088/0004-637X/788/1/73}

\end{thebibliography}
\bibliographystyle{aasjournalv7}
\end{CJK*}
\end{document}